%% file: main.tex
\documentclass[a4paper, amsfonts, amssymb, amsmath, reprint, showkeys, twoside, superscriptaddress, aps]{revtex4-1}
\input{preamble}

\begin{document}

\title{Quasi-Probabilistic Readout Correction of Mid-Circuit Measurements for Adaptive Feedback via Measurement Randomized Compiling}

\author{Akel Hashim}
    \thanks{These authors contributed equally to this work. Correspondence should be addressed to \href{mailto:ahashim@berkeley.edu}{ahashim@berkeley.edu}.}
    \affiliation{Quantum Nanoelectronics Laboratory, Department of Physics, University of California at Berkeley, Berkeley, CA 94720, USA}
    \affiliation{Applied Math and Computational Research Division, Lawrence Berkeley National Lab, Berkeley, CA 94720, USA}
\author{Arnaud Carignan-Dugas}
    \thanks{These authors contributed equally to this work. Correspondence should be addressed to \href{mailto:ahashim@berkeley.edu}{ahashim@berkeley.edu}.}
    \affiliation{Keysight Technologies Canada, Kanata, ON K2K 2W5, Canada}
\author{Larry Chen}
    \affiliation{Quantum Nanoelectronics Laboratory, Department of Physics, University of California at Berkeley, Berkeley, CA 94720, USA}
\author{Christian Jünger}
    \affiliation{Quantum Nanoelectronics Laboratory, Department of Physics, University of California at Berkeley, Berkeley, CA 94720, USA}
    \affiliation{Applied Math and Computational Research Division, Lawrence Berkeley National Lab, Berkeley, CA 94720, USA}
\author{Neelay Fruitwala}
\author{Yilun Xu}
\author{Gang Huang}
    \affiliation{Accelerator Technology and Applied Physics Division, Lawrence Berkeley National Lab, Berkeley, CA 94720, USA}
\author{Joel J. Wallman}
    \affiliation{Keysight Technologies Canada, Kanata, ON K2K 2W5, Canada}
\author{Irfan Siddiqi}
    \affiliation{Quantum Nanoelectronics Laboratory, Department of Physics, University of California at Berkeley, Berkeley, CA 94720, USA}
    \affiliation{Applied Math and Computational Research Division, Lawrence Berkeley National Lab, Berkeley, CA 94720, USA}
    \affiliation{Materials Sciences Division, Lawrence Berkeley National Lab, Berkeley, CA 94720, USA}

\date{\today}

\begin{abstract}
    Quantum measurements are a fundamental component of quantum computing. However, on modern-day quantum computers, measurements can be more error prone than quantum gates, and are susceptible to non-unital errors as well as non-local correlations due to measurement crosstalk. While readout errors can be mitigated in post-processing, it is inefficient in the number of qubits due to a combinatorially-large number of possible states that need to be characterized. In this work, we show that measurement errors can be tailored into a simple stochastic error model using randomized compiling, enabling the efficient mitigation of readout errors via quasi-probability distributions reconstructed from the measurement of a single preparation state in an exponentially large confusion matrix. We demonstrate the scalability and power of this approach by correcting readout errors without matrix inversion on a large number of different preparation states applied to a register of eight superconducting transmon qubits. Moreover, we show that this method can be extended to mid-circuit measurements used for active feedback via quasi-probabilistic error cancellation, and demonstrate the correction of measurement errors on an ancilla qubit used to detect and actively correct bit-flip errors on an entangled memory qubit. Our approach enables the correction of readout errors on large numbers of qubits, and offers a strategy for correcting readout errors in adaptive circuits in which the results of mid-circuit measurements are used to perform conditional operations on non-local qubits in real time.
\end{abstract}


\maketitle

\input{sections/1_intro}
\input{sections/2_rc}

\input{sections/3_qp}
\input{sections/4_mcm}
\input{sections/5_conclusions}
\input{sections/acknowledgements.tex}

\bibliography{bibliography}

\clearpage

\appendix
\input{sections/appendices.tex}

\end{document}

%% file: preamble.tex
\usepackage{amsthm}
\usepackage{amssymb}
\usepackage[english]{babel}
\usepackage[utf8]{inputenc}
\usepackage[colorinlistoftodos, color=green!40, prependcaption]{todonotes}
\usepackage{mathtools}
\usepackage{physics}
\usepackage{xcolor}
\usepackage{graphicx}
\usepackage[left=23mm,right=13mm,top=35mm,columnsep=15pt]{geometry} 
\usepackage{adjustbox}
\usepackage{placeins}
\usepackage[T1]{fontenc}
\usepackage{lipsum}
\usepackage{csquotes}
\usepackage{tikz}
\usepackage{bm}
\usepackage{braket}
\usepackage{multirow}
\usepackage{qcircuit}
\usepackage{siunitx}
\usepackage{txfonts}

\usepackage[font=small,labelfont=bf,
   justification=Justified,
   format=plain]{caption}
\usepackage{subcaption}

\usepackage{hyperref}
\hypersetup{
    colorlinks=true,
    linkcolor=blue,
    urlcolor=blue,
    citecolor=blue
    }
\usepackage[capitalise]{cleveref}

\newcommand{\eq}{Eq.~}
\newcommand{\fig}{Fig.~}

\newcommand{\bs}[1]{{\underline{#1}}_{AB}}
\newcommand{\0}[0]{\underline{0}}
\newcommand{\mbb}[1]{\mathbb{#1}}
\newcommand{\bsmarg}[2]{{\underline{#1}}_{#2}}

%% file: sections/1_intro.tex
\section{Introduction}\label{sec:intro}

Measurement plays a foundational role in quantum mechanics. It is the means by which we learn properties of quantum systems, and is fundamentally linked with the collapse of quantum wavefunctions. Measurement is also essential to quantum computing. In gate-based quantum computing, measurement is needed to translate quantum bits (qubits) to classical bits at the end of a computation, it is the central component in teleportation-based protocols \cite{gottesman1999demonstrating, chou2018deterministic} and measurement-based quantum computing \cite{raussendorf2001one, briegel2009measurement}, it can be utilized to generate long-range entanglement in constant depth via adaptive quantum circuits \cite{foss2023experimental, hashim2024efficient}, and it is necessary for syndrome extraction in quantum error correction \cite{shor1995scheme, knill1997theory, chiaverini2004realization, ofek2016extending, rosenblum2018fault, livingston2022experimental}. However, measurements are inherently noisy, and the nature of errors can depend not only on the quantum state prior to measurement, but can also contextually depend on the state of other qubits. Moreover, measurements are often slower and more error prone than the unitary gates used to prepare quantum states, which places limits on the speed and fidelity with which they can be used to perform real-time corrections in the middle of quantum circuits.

A common assumption in quantum computing is that readout errors are purely probabilistic --- that is, for a given projective measurement of some finite duration, a qubit has a defined probability of experiencing a bit flip during readout. However, this assumption is often violated in systems with multiplexed readout in which measurement crosstalk can cause context-dependent coherent and correlated readout errors \cite{blumoff2016implementing, heinsoo2018rapid, chen2019detector, maciejewski2020mitigation}. Moreover, the probability of a bit-flip for a qubit in an excited state can vastly differ from the probability of a bit-flip while sitting in the ground state due to non-unital processes such as energy relaxation and $T_1$ decay \cite{elder2020high}, leading to context-dependent errors which depend on the state of a qubit prior to readout. However, by twirling a process over a unitary 1-design, one can effectively \emph{design} stochastic channels \cite{graydon2022designing}. One such strategy for designing stochastic channels is randomized compiling (RC) \cite{wallman2016noise, hashim2021randomized}, which is a robust and efficient method for tailoring arbitrary Markovian errors into Pauli channels in gate-based quantum computing. While RC was originally designed for tailoring gate noise, it can be adapted to tailor measurement noise \cite{beale2023randomized}, and has been previously shown to reduce worst-case error rates in state-preparation and measurement (SPAM) \cite{hashim2023benchmarking, mclaren2023stochastic}. In this work, we experimentally deploy RC for quantum measurements on an eight-qubit superconducting quantum processor (see \fig\ref{fig:fig1}a). We show that, under measurement RC (MRC), quantum measurement noise can be accurately described by a stochastic error model in which the probability of a bit-flip for any given qubit is independent of the preparation state or the state of other qubits on the quantum processor, thus enforcing common pre-existing assumptions about the stochasticity of measurement errors.

Because measurements translate quantum bits to classical bits, errors in terminating measurements can be mitigated classically via post-processing. One approach requires preparing and measuring all possible combinations of input basis states for $n$ qubits, from which one can construct a \emph{confusion matrix} of the measured results. Applying the inverse of the confusion matrix on the resulting outcome distribution often mitigates measurement errors. This strategy is limited to a subset of measurement errors, since the confusion matrix does not capture the effect of coherent measurement errors on quantum superpositions. Another limitation of this approach is its poor scalability: the size of this matrix grows exponentially in the number of qubits, making both the characterization and inversion steps intractable for large qubit numbers. As a result, experimentalists often resort to performing \emph{local} readout correction \cite{bravyi2021mitigating}, in which an individual confusion matrix is measured and inverted for each qubit. While this can correct individual readout errors, it cannot correct correlated bit-flips. Alternative strategies for improving qubit readout include encoding qubits in a repetition code prior to readout \cite{gunther2021improving, hicks2022active}, which comes at the cost of additional ancillae qubits, or correcting readout errors based on the results of detector tomography \cite{chen2019detector, maciejewski2020mitigation}. By tailoring noise in measurements into a stochastic bit-flip channel, we show that it is possible to correct readout errors for any input state without matrix inversion, ancillae qubits, or full tomography. To do so, it is sufficient to characterize readout errors on a single input state (e.g., $\ket{0^{\otimes n}}$ for $n$ qubits) under MRC, from which a quasi-probability distribution can be constructed. Readout correction is then performed by inverting the quasi-probability distribution on the measured bit-string results. We compare full and local readout correction to our quasi-probabilistic protocol for a large number of structured and random input states on eight qubits, and show that our protocol improves the results in over 90\% of the circuits. Moreover, we show that this scheme extends to mid-circuit measurements (MCMs), and demonstrate the mitigation of readout errors used to perform real-time feedback to correct for bit-flip errors on an entangled qubit.

%% file: sections/2_rc.tex
\section{Randomized Compiling for Measurements}\label{sec:rc}

\begin{figure*}[ht]
    \centering
    \includegraphics[width=2\columnwidth]{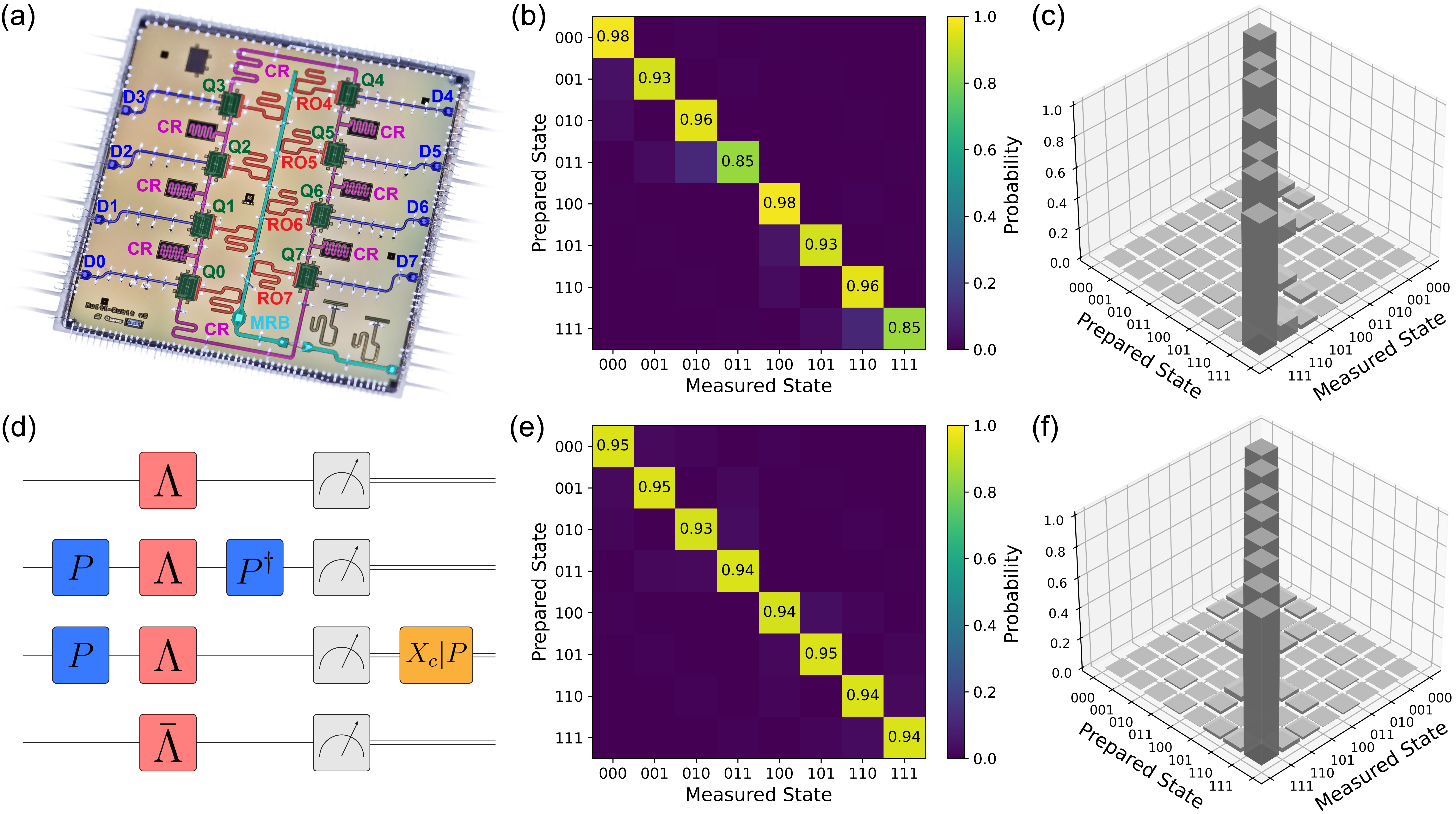}
    \caption{\textbf{Randomized Compiling for Measurements.}
    \textbf{(a)} 8-qubit superconducting transmon processor. Qubits are labeled in green, individual drive lines are labeled in blue, individual readout resonators (RO) are labeled in red, and the multiplexed readout bus (MRB) is labeled in cyan. The qubits are coupled to nearest neighbors in a ring geometry via coupling resonators (CR, purple). 
    \textbf{(b)} Full confusion matrix measured for three qubits (Q3, Q4, Q5). Strong state-dependent errors are observed. For example, when $\ket{011}$ or $\ket{111}$ is prepared, $\ket{010}$ and $\ket{110}$ are measured $\sim$11\% of the time, respectively.
    \textbf{(c)} The confusion matrix in (b) minus the identity matrix. Context-dependent errors --- such as errors that depend on the state of a qubit prior to measurement --- appear as an asymmetry in the off-diagonal terms of the confusion matrix.
    \textbf{(d)} [Top] We can model the error in a measurement by a process matrix $\Lambda$ preceding the measurement. [Second] In theory, it is possible to twirl this process matrix via Pauli twirling, $\Lambda \mapsto P^\dagger \Lambda P$. [Third] However, because this is a process matrix for a (non-unitary) measurement, the inversion operators must be implemented as classical bit-flips $X_c$ conditioned on which Pauli $P$ was sampled before the measurement. [Bottom] By averaging this measurement many times over the full Pauli group, we obtain a twirled error process $\bar{\Lambda} = \tfrac{1}{4} \sum_{P \in \{I, X, Y, Z\}} P^\dagger \Lambda P$, in which measurement errors have been reduced to a stochastic bit-flip channel. 
    \textbf{(e)} Full three-qubit confusion matrix measured using the scheme presented in (d) for qubits Q3, Q4, Q5. We observe that the diagonal entries of the confusion matrix are all approximately equal in magnitude, showing that we have eliminated state-dependent readout errors.
    \textbf{(f)} The confusion matrix in (e) minus the identity matrix.  All error probabilities in the off-diagonal elements are (approximately) symmetric along the diagonal. This indicates that, under MRC, the probability of a bit flip for a given qubit is the same for all states.
    }
    \label{fig:fig1}
\end{figure*}

Generalized measurements of quantum states are described by positive-operator valued measures (POVMs), which are set of positive semi-definite Hermitian matrices $\{E_i\}$ in $d$-dimensional Hilbert space $\mathcal{H}_d$ that obey the completeness relation:
\begin{equation}\label{eq:completeness_rel}
    \sum_i E_i = \mathbb{I}~.
\end{equation}
The probability of measuring an outcome $i$ given a state $\rho$ is governed by Born's rule,
\begin{equation}
    p(i|\rho) = \Tr[E_i\rho]~.
\end{equation}
For a given system containing $n$ qubits, the POVM set corresponding to computational basis measurements contains $2^n$ elements, $\{E_i\}_{i=1}^{2^n}$, with each element indexed by an $n$-qubit bit string $i$. For example, for a single qubit the POVM set is $\{E_0, E_1\}$, for two qubits the POVM set is $\{E_{00}, E_{01}, E_{10}, E_{11}\}$, etc. By preparing a system of $n$ qubits in all $2^n$ possible combinations of basis states, represented by the set of input states $\{\rho_j\}$, and measuring the resulting POVMs $\{E_i\}$ for each basis state, one can construct a $2^n \times 2^n$ \textit{confusion matrix} $\mathcal{M} = \langle\braket{ \{E_i\} | \{\rho_j\} }\rangle$ whose elements
\begin{equation}\label{eq:cmat_elements}
    \mathcal{M}_{ij} = \Tr[E_i \rho_j]
\end{equation}
represent the probability $p(i|j)$ of measuring the outcome $E_i$ given an input state $\rho_j$, where the double-bra (-ket) notation $\bra{\langle \cdot}$ ($\ket{\cdot \rangle}$) denotes the \emph{vectorization} of the POVM $E_i$ (initial state $\rho_j$) into a $1 \times d^2$ row vector ($d^2 \times 1$ column vector). Classically, the confusion matrix is sufficient to predict the probability distribution of an outcome given any input. However, in quantum computing, while the $d \times d$ confusion matrix is an experimentally well-defined object (see \fig\ref{fig:fig1}b -- f), it generally only provides one part of the picture. Indeed, while \eq\ref{eq:cmat_elements} correctly describes the probabilities to observe the outcome $i$ given the computational state $j$, it generally does not correctly prescribe the probability distribution expected for a quantum state involving quantum superpositions \cite{beale2023randomized}. Fortunately, there is a way to compile quantum circuits such that, statistically, confusion matrices fully prescribe measurement errors as in the classical case. Such method, which we call measurement randomized compiling (MRC), was introduced in \cite{beale2023randomized} and is described further below.

Let us consider the scenario where the measurement error is such that given an $n$-qubit confusion matrix $\mathcal{M}$ and an ideal probability distribution $p$, the effect of measurement noise on the ideal outcomes produces a noisy probability distribution $q = \mathcal{M} p$. In such case, correcting the effect of measurement noise on a probability distribution reduces to inverting $\mathcal{M}$ given a measured distribution $q$:
\begin{equation}\label{eq:readout_correction}
    p = \mathcal{M}^{-1} q ~.
\end{equation}
If $\mathcal{M}$ is known and if it correctly models measurement errors, then in theory one can correct the effect of measurement errors affecting the outcome of any quantum circuit. However, because $\mathcal{M}$ scales exponentially in the number of qubits $n$, in practice it is not feasible to construct a full $n$-qubit confusion matrix, nor is it always necessary if one can make reasonable assumptions about the locality and nature of correlated measurement noise. An alternative strategy is to assume that readout errors are uncorrelated and that measurement noise can be modeled as a tensor product of confusion matrices. In this case, it is sufficient to reconstruct the individual confusion matrix for each qubit, such that $\mathcal{M}$ is given as 
\begin{equation}\label{eq:cmat_tensor}
    \mathcal{M} = \prod_{i=1}^n \otimes \mathcal{M}_i ~,
\end{equation}
where $\mathcal{M}_i$ is the confusion matrix for the $i$th qubit. Now, the inversion process (\eq\ref{eq:readout_correction}) only corrects readout errors on each qubit individually, but cannot account for any correlated readout errors. 

While it is often assumed that readout errors are probabilistic and locally independent, in which case measuring individual confusion matrices for each qubit would be sufficient to correct all readout errors, in practice this is not the case. For example, in \fig\ref{fig:fig1}b, we plot the full confusion matrix for three qubits (Q3, Q4, and Q5; see \fig\ref{fig:fig1}a). We observe that for most preparation states, the combined readout fidelity is between 93\% -- 98\%. However, for $\ket{011}$ and $\ket{111}$, we observe poor readout fidelities; these probabilities are inconsistent with an assumption of independent bit-flip rates. Such errors could be in part due to the fact that the readout frequencies of Q3, Q4, and Q5 are close in frequency (see Appendix \ref{sec:appendix}) resulting from fabrication inaccuracies, leading to readout crosstalk \cite{blumoff2016implementing, chen2019detector}, which can result in context-dependent readout errors in which the error on a qubit depends on the state of \emph{another} qubit. Moreover, even for the preparation states with higher readout fidelities, we generally observe that excited states have worse readout fidelity than ground states. This is due to non-unital errors such as $T_1$ decay, which results in state-dependent readout errors, placing fundamental limits on excited state fidelities for a given readout time \cite{elder2020high}.

Fortunately, there is a way to simplify the situation drastically by statistically tailoring the measurement error into a probability distribution over bit-flips that remains independent of the measured state. The idea is to twirl the readout noise using compiling methods such as MRC \cite{beale2023randomized}. To understand the principles behind MRC, let us express a noisy measurement $\langle\bra{\tilde{E_i}}$ as an ideal measurement $\langle\bra{E_i}$ preceded by a process matrix $\Lambda$ which captures all measurement errors: $\langle\bra{\tilde{E_i}} = \langle\bra{E_i} \Lambda$ (see \fig\ref{fig:fig1}d). The goal of MRC is to twirl $\Lambda$ into diagonal Pauli channels, i.e., $\Lambda \mapsto 4^{-n}\sum_{P \in \mathbb{P}_n} P^\dagger \Lambda P$, where $\mathbb{P}_n = \{I, X, Y, Z\}^{\otimes n}$ is the $n$-qubit Pauli group. However, in reality, readout errors occur concurrently with measurement; therefore, we cannot simply conjugate $\Lambda$ by Pauli gates. Rather, to twirl $\Lambda$ we compile random Paulis into the final cycle of single-qubit gates before measurement and perform classical bit-flips on the measured results conditional on the inserted Pauli for each qubit. For example, if $I$ or $Z$ is inserted, these will not change the results of measurements in the computational basis; however, if $X$ or $Y$ is sampled, these will flip the qubit state prior to measurement, necessitating classical bit-flips after measurement. By repeating this process many ($K$) times and recording the combined distribution of all results, we obtain an effective Pauli-twirled process matrix
\begin{equation}\label{eq:sample_avg}
    \bar{\Lambda}_K = \frac{1}{K} \sum_{\substack{ i=1 \\ P_i \in_R \mathbb{P}_n}}^{K} P_i^\dagger \Lambda P_i~,
\end{equation}
where $R$ denotes that $P_i$ is chosen at random from the $n$-qubit Pauli group $\mathbb{P}_n$ each time. $\bar{\Lambda}_K$ 
in \cref{eq:sample_avg} is a sample average, and it converges quickly to the true average 
$\bar{\Lambda}_\infty =: 4^{-n}\sum_{P \in \mathbb{P}_n} P^\dagger \Lambda P$, as shown in the theoretical analysis of RC \cite{wallman2016noise, Wallman2018, Winick2022}. More importantly, the convergence of the sample average to the true average is almost independent of the system size (just like the required sample size of a poll is almost independent of the population size); this property is what makes RC applicable to any  system size (i.e., number of qubits or Hilbert space dimension) \cite{goss2023extending}. To demonstrate that MRC scales to more qubits in practice, we repeat the same analysis on all eight qubits on our quantum processor (see Appendix \ref{sec:8q_mrc_qprc}), showing indeed that MRC tailors readout noise equally well on eight qubits as 
on three.

The appeal of RC is that the true average error $\bar{\Lambda}_\infty$ can provably be expressed 
as a probabilistic mixture of Pauli gates (also known as a Pauli stochastic channel) \cite{wallman2016noise,Wallman2018,Winick2022}:
\begin{align}
    \bar{\Lambda}_\infty[\rho] = \sum_{P \in \mathbb{P}_n} p(P) P^\dagger \rho  P ~, 
\end{align}
where $p(P)$ is the probability of the Pauli error $P \in \mathbb{P}_n$. In the case of measurement, 
$Z$ have no effect on computational basis states, and $Y$ has the same effect as $X$. Therefore, we get a classical stochastic error channel of the form:
\begin{align}\label{eq:twirled_meas}
    \bar{\Lambda}_K[\rho] \approx \bar{\Lambda}_\infty[\rho] = \sum_{x \in \mathbb{Z}_2^{\otimes n}} p_x X^x\rho X^x ~,
\end{align}
where $x \in \mathbb{Z}_2^{\otimes n}$ is the set of classical $n$-bit strings, $\{p_x\}_{x\in \mathbb{Z}_2^{\otimes n}}$ is a probability distribution over bit-flips $X^x$, and where $X^x$ is short for $X^{x_1} X^{x_2} \cdots X^{x_n}$.

In \fig\ref{fig:fig1}e, we plot the full confusion matrix for qubits Q3, Q4, and Q5 reconstructed using MRC with $K=100$ randomizations \cite{hashim2023benchmarking}. We observe that the diagonal readout fidelities $p(i|i)$ are all approximately equal, and the off-diagonal probabilities $p(i|j) \; \forall \; j \ne i$ are approximately symmetric along the diagonal, suggesting that we have eliminated state- and context-dependent readout errors due to $T_1$ decay and readout crosstalk. This provides experimental evidence that under MRC, we can describe readout errors as a purely stochastic process in which the probability of a bit-flip for any given qubit is independent of the preparation state (see also Appendix \ref{sec:8q_mrc_qprc}).

It should be noted that a similar method to MRC was introduced in \cite{smith2021qubit, hicks2021readout} by inserting bit-flips prior to measurement. However, bit-flip averaging does not provide a complete twirl of the readout noise, as phase randomization is also necessary in order to describe readout errors as purely stochastic. For example, suppose a qubit is in the $\ket{i+}$ state prior to measurement; here, a coherent-$X$ error during measurement \cite{chen2019detector} will result in an incorrect result distribution. However, by randomly inserting Pauli-$Z$ gates prior to measurement, the impact of the coherent-$X$ error will be averaged away on the ensemble level \footnote{While Pauli-$Z$ gates can be implemented entirely via phase shifts in the following pulse, this will have no effect if the gate directly precedes a measurement. Instead, we implement Pauli-$Z$ gates using the $ZXZXZ$ decomposition of the gate, such that physical pulses are always played prior to measurement.}. Finally, it is worth noting that because readout errors are state-independent under MRC, the effect of readout errors on Pauli expectation values can be efficiently corrected by re-scaling by the average readout fidelity \cite{van2022model}.

%% file: sections/3_qp.tex
\section{Quasi-Probabilistic Readout Correction}\label{sec:qp}

\begin{figure*}[ht]
    \centering
    \includegraphics[width=2\columnwidth]{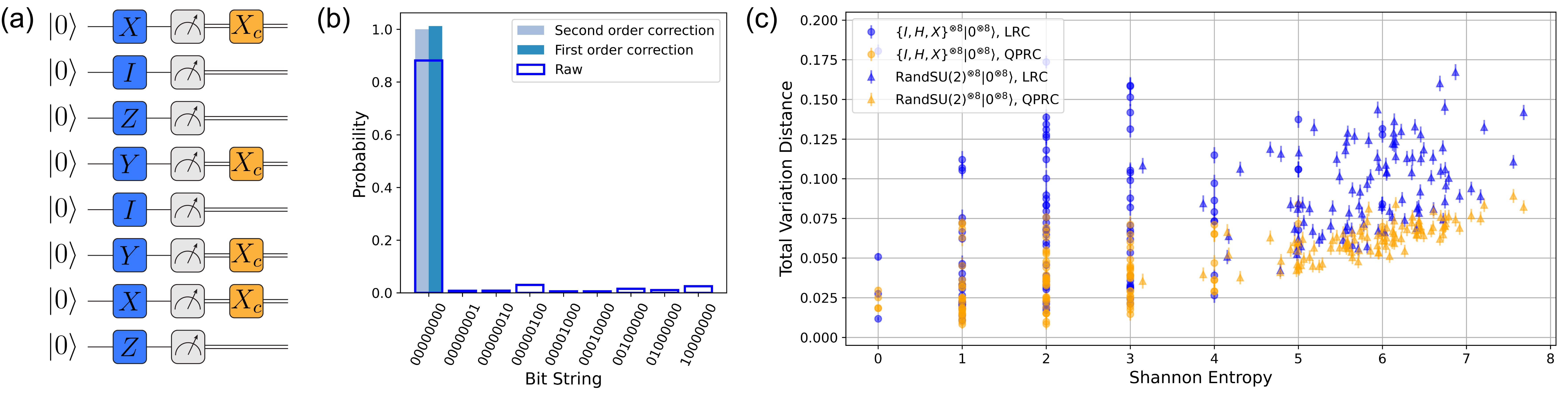}
    \caption{\textbf{Quasi-Probabilistic Readout Correction.} 
    \textbf{(a)} Readout characterization. The probability of bit-flip errors during readout can be characterized by preparing a single $n$-qubit input state (e.g., $\ket{00000000}$ for eight qubits) and measuring the resultant states under RC. To do so, a randomly-sampled $n$-qubit Pauli operator should be inserted before measurement (e.g., $X \otimes I \otimes Z \otimes Y \otimes I \otimes Y \otimes X \otimes Z$); after measurement, a classical bit-flip $X_c$ should be applied to any qubit in which an $X$ or $Y$ gate was applied before measurement. This process should be repeated many times for many different randomly-sampled Pauli operators.
    \textbf{(b)} Results for the protocol in (a) applied to eight qubits using $K = 100$ different randomizations is plotted with a blue outline. We observe single-qubit bit-flip errors on many qubits. For example, while $\ket{00000000}$ is measured over 88\% of the time, we observe that Q5 has a $\sim$3\% chance of experiencing a bit-flip error during readout. This distribution can be used to perform quasi-probabilistic readout correction on any 8-qubit circuit in which the readout is performed using MRC. The first- and second-order corrections performed on the distribution itself are plotted in blue and light blue, respectively. The first-order correction has a probability greater than 1.0 for $\ket{00000000}$ (and small negative counts for other bit strings); however, the second-order correction reconstructs a distribution in which only the all-zero state remains. (Only the bit strings with significant counts are displayed for clarity.)
    \textbf{(c)} Readout-corrected single-qubit circuits. Structured circuits were generated by applying gates randomly sampled from $\{I, H, X\}$ to each qubit (circular data points), and random circuits were generated by applying a random $\mathsf{SU(2)}$ gate independently to each qubit (triangular data points). Each circuit was performed with and without MRC, denoted by the orange and blue/purple data points, respectively. For circuits without MRC, we apply local readout correction (LRC; blue) using confusion matrices measured for each qubit, or full readout correction (FRC), using a full 8-qubit confusion matrix. For the circuits with MRC, we apply the QPRC protocol using the distribution measured in (b). We plot the TVD of the experimental results with the ideal results as a function of the Shannon entropy $S$ of the ideal result. $S=0$ corresponds to a singular distribution, and $S=8$ is the uniform distribution for 8 qubits; in general, the larger the entropy, the more uniform the distribution. We observe that the TVD of the MRC + QPRC results are lower than the results obtained with either LRC or FRC in over 90\% of the circuits; thus, MRC + QPRC broadly outperforms both LRC and FRC. (Error bars for the TVD are on the order of the size of the markers.)
    }
    \label{fig:fig3}
\end{figure*}

As observed in the previous section, applying RC to quantum measurements effectively tailors the measurement error channel $\Lambda$ into a classical stochastic error channel. This has a few ramifications: firstly, the effective error channel $\bar{\Lambda}_K$ can be fully described by its corresponding probability distribution, and each probability $p_x$ can be estimated up to $1/\sqrt{N_{\rm shots}}$ simply by looking at the output distribution resulting from sending a single computational basis input to the randomly compiled measurement channel. In other words, $\bar{\Lambda}_K$ can be approximately described with $\mathcal{O}(N_{\rm shots})$ floating-point numbers, and each number has a precision of $1/\sqrt{N_{\rm shots}}$ \footnote{The accuracy of the estimate is also limited by state preparation errors and single-qubit gate errors, but these tend to be low compared to measurement errors.}.
Secondly, the effective measurement error $\bar{\Lambda}_K$ can be inverted by applying a linear operation on the noisy output distribution. The exact inversion can quickly become unscalable to describe, but since the probabilities appearing in $\bar{\Lambda}_K$ are already estimated with $1/\sqrt{N_{\rm shots}}$ precision, an approximation should suffice. Fortunately, there exist standard quasi-probabilistic correction techniques that provide different orders of approximation of the inverse of $\bar{\Lambda}_K$ \cite{ying2017efficient,Temme2017mitigation,endo2018mitigation,Kandala_2019,ferracin2022efficiently}. The first-order approximation is described using $\mathcal{O}(N_{\rm shots})$ floating-point numbers, and in general the $i$th order approximation is described using $\mathcal{O}(N_{\rm shots}^i)$ floating-points numbers. We proceed with describing the quasi-probabilistic readout correction (QPRC) protocol below.

Because measurement errors under MRC can be described by a stochastic bit-flip channel which is independent of the input state, it is sufficient to characterize the probability of bit-flips on $n$ qubits using a single preparation state. For simplicity, we choose to characterize measurement errors on $\ket{0^{\otimes n}}$ using the MRC protocol. For example, in \fig\ref{fig:fig3}a we depict a single cycle of Paulis applied to the all-zero state on eight qubits; if $X$ or $Y$ is applied before measurement, then a classical bit-flip $X_c$ is applied in post-processing. This process should be repeated many ($K$) times to construct a twirled measurement channel (\eq\ref{eq:twirled_meas}). In \fig\ref{fig:fig3}b, we plot the results of the characterization procedure in blue using $K=100$ randomizations. We observe that the all-zero state is measured over 88\% of the time, with the remaining $\sim$12\% distributed over various single-qubit bit-flip channels. 

To model readout noise under MRC, let us denote linear combinations of bit-strings as $\sum_{x} c_x x$, where $c_x \in \mbb{R}$ are scalar coefficients and $x \in \mbb {Z}_n$ are bit-strings.
Given an error probability distribution $p = \sum_x p_x x$ and an ideal outcome distribution $a = \sum_y a_y y$, we can express the resulting noisy outcome distribution 
$b = \sum_z b_z z$ as
\begin{align}\label{eq:b}
    b &= p \oplus a ~, \notag \\
    &=\left(\sum_x p_x x \right) \oplus \left(\sum_y a_y y\right) ~,
\end{align}
where $x \oplus y$ is the \emph{bitwise} modulo 2 sum of the $x$ and $y$ bit-strings. 
In other words, the probability of observing the outcome $z$ given a noisy measurement is
\begin{equation}
    b_z  = \sum_{x \oplus y = z} p_x a_y ~.
\end{equation}

Now, to invert the effect of readout errors on the distribution $a$, we perform Probabilistic Error Cancellation (PEC) \cite{Temme2017mitigation, endo2018mitigation} for measurement. That is, we construct a quasi-probability distribution $q$ which is an approximate inverse of $p$. If we denote the all-zero bit-string as $\0$ (we underline bit-strings to distinguish them from scalar coefficients), the goal is to construct $q$ such that $q \oplus p \approx \0$ (notice that $\0$ is the identity w.r.t.~to the $\oplus$ operator). Indeed, suppose that $q \oplus p =\0$, 
then it follows from \eq\ref{eq:b} that applying $q$ to $b$ yields $q \oplus b = q \oplus p \oplus a = \0 \oplus a =a$.
A first-order construction of $q$ can be expressed as follows:
\begin{equation}\label{eq:qp_dist}
    q = \frac{1}{2p_{\0}-1} \left( p_{\0} \0 - \sum_{x \ne \0} p_x x \right) ~.
\end{equation}
Indeed, applying $q$ to $p$ yields
\begin{align}
    p' = q \oplus p &= \frac{1}{2p_{\0}-1} \left( p_{\0} {\0} - \sum_{x \ne {\0}} p_x x \right) \oplus \left( p_{\0} {\0} + \sum_{x \ne {\0}} p_x x  \right) ~, \notag \\
              &=  \frac{p_{\0}^2}{2p_{\0}-1} {\0} - \frac{1}{2p_{\0}-1}\left( \sum_{x \ne {\0}} p_x x  \right)^2 ~,
\end{align}
which is approximately equal to the identity up to second-order. To put it simply, the error amplitude goes from $1-p_{{\0}}$ to $(1-p_{{\0}})^2/(2p_{{\0}}-1)$.
However, the inverse operation can be improved further. Consider the family of quasi-probability distributions:
\begin{equation}\label{eq:qp_dist_k}
    q^{(k)} = \frac{p_{{\0}}^{2k-1}}{p_{\0}^{2k}-(1-p_{\0})^{2k}} \left( {\0}+\sum_{j=1}^{2k-1} \left(\frac{-1}{p_{{\0}}} \right)^{j} \left(\sum_{x \ne {\0}} p_x x\right)^j \right) ~,
\end{equation}
for $k \in \mathbb{N}_+$. Notice that $q^{(1)} = q$ from 
\cref{eq:qp_dist}. Applying $q^{(k)}$ to $p$ yields:
\begin{align}\label{eq:qk}
    q^{(k)} \oplus p = \frac{p_{{\0}}^{2k}}{p_{\0}^{2k}-(1-p_{\0})^{2k}} {\0} - \frac{1}{p_{\0}^{2k}-(1-p_{\0})^{2k}} \left(\sum_{x \ne {\0}} p_x x\right)^{2k }~.
\end{align}
Thus, in the generalized case, the error amplitude goes from $1-p_{{\0}}$ to $\frac{1}{1 - (p_{{\0}}/(1 - p_{\0}))^{2k}}$. This strategy is expected to be effective as long as $p_{{\0}}> 1/2$, which is the turning point for which the coefficient in front of $\0$ in \eq\ref{eq:qk} does not converge to 1 by increasing $k$. In principle, the constraint $p_{{\0}}> 1/2$ limits the scaling of this method. However, in Appendix \ref{sec:generalized_inversion}, we generalize the above quasi-probability inversion method, allowing it to scale for large systems where the total error probability $1 - p_{\0}$ reaches well above $1/2$. Of course, like any mitigation method, there is a limit to its scalability \cite{Takagi_2022}. However, given the current observations, since it does not involve any error propagation through a circuit, it is expected to apply well to outcome distributions \emph{marginalized} over tens of physical qubits (given the same error rates).

In practice, to correct readout errors on any noisy experimental probability distribution $b$ which has been measured using MRC, we sum over all corrected results in which the counts for each experimental results $x$ have been redistributed according to $q^{(k)}$:
\begin{align}\label{eq:qp_rcorr}
    b^{(k)} &= \sum_x \left( b_x x \oplus q^{(k)} \right) ~,
\end{align}
where the readout corrected distribution $b^{(k)}$ is the union over all of the redistributed counts $b_x x \oplus q^{(k)}$. To demonstrate that our procedure corrects readout errors, we perform a first- (i.e., using $q^{(1)}=q$) and second-order (i.e., using $q^{(2)}$) correction on the characterized probability distribution in \fig\ref{fig:fig3}b that is used to construct the quasi-probability distribution. We observe that the first-order correction redistributes most of the results to ${\0}$, but that $p^{(1)}_{\0}$ is slightly greater than 1.0. Because the corrected distribution is itself a quasi-probability distribution that has been normalized to preserve the total probability, ${\0}$ has a quasi-probability $p^{(1)}_{{\0}}$ greater than 1.0 to account for the negative quasi-probabilities in the other states (not shown). It is reasonable for the first-order correction to have negative probabilities, because the quasi-probability distribution is only an \emph{approximate} inverse distribution; thus, small residual biases can remain after the first-order correction. However, after performing a second-order correction on the characterized distribution, we find that $p^{(2)}_{{\0}} = 1.0$, as we would expect if all readout errors were corrected. This process highlights the fact that readout correction (or, more generally, error mitigation strategies) can introduce non-physical outcomes into the results of experiments. For example, if one wants to preserve the total probability of a process, then the small residual negative values that remain after the quasi-probabilistic readout correction should be preserved, which equates to enforcing trace-preservation (TP). However, negative probabilities violate complete-positivity (CP), and these values could be reasonably set to zero depending on the nature of the final computation. Therefore, one cannot always enforce both CP and TP on the outcomes of error corrected results, and the choice of which to preserve is up to the experimenter. 

To demonstrate the efficacy of our protocol on a wide variety of input states, we perform a second-order correction (i.e., using $q^{(2)}$) on 200 different eight-qubit input states prepared using a single cycle of gates, shown in \fig\ref{fig:fig3}c. For half of the circuits, we sample gates from $\{ I, H, X \}$ at random for each qubit, and for the other half of the circuits we sample random $\mathsf{SU(2)}$ gates for each qubit independently. To compute the accuracy of the readout corrected results, we compute the total variation distance (TVD) between the experimental distribution $b$ and the ideal distribution $a$,
\begin{equation}\label{eq:tvd}
    D_\text{TV}(b,a) = \frac{1}{2} \sum_x |b_x - a_x| ~,
\end{equation}
plotted as a function of the Shannon entropy of the ideal results,
\begin{equation}\label{eq:entropy}
    S = -\sum_x a_x \log_2 (a_x) ~.
\end{equation}
A lower value for $D_\text{TV}$ means that the results are closer to the ideal distribution. For the Shannon entropy, $S=0$ corresponds to a probability distribution that is peaked around a single value, and for 8 qubits $S=8$ is the uniform distribution; in general, the higher the entropy, the closer to a uniform distribution. 
We compare the results of our QPRC protocol to results obtained using full readout correction (FRC) using matrix inversion of the full confusion matrix, which is not scalable, and local readout correction (LRC), which is scalable. We find that while FRC and LRC perform approximately equivalently over all circuits (i.e., their average TVDs are equivalent), our protocol produces better results (i.e., has a lower TVD) in over 90\% of the circuits compared to both FRC and LRC.
Additionally, we observed a positive linear correlation between the TVD of the corrected results and the entropy of the ideal results, with better performance at lower entropy. This can be explained by the fact that for higher entropy, the approximate correction has to be applied to more outputs, meaning that the systematic error in the approximated inverse is applied more often.

It should be noted that QPRC makes no distinction between the different sources of physical errors that lead to readout errors (e.g., $T_1$ decay, misclassification due to low measurement signal-to-noise, etc.). Therefore, it can correct different readout errors equally well, as under MRC they all appear as stochastic bit flips. However, like other methods for performing readout correction, periodic re-characterization of the readout errors under MRC is necessary for accurate readout correction via QPRC. While MRC is robust to drift and, for example, fluctuations in qubit $T_1$ times (which would cause excited state readout fidelities to also fluctuate in time), QPRC requires an accurate characterization of the error probability distribution $p$ in order to construct the inverse quasi-probability distribution $q$. Thus, it is recommended that one re-characterize the bit-flip error rates under MRC periodically, depending on how often one expects the system to drift or $T_1$ times to fluctuate. 

%% file: sections/4_mcm.tex
\section{Readout Correction for Mid-Circuit Measurements}\label{sec:mcm}

\begin{figure*}[t]
    \centering
    \includegraphics[width=2\columnwidth]{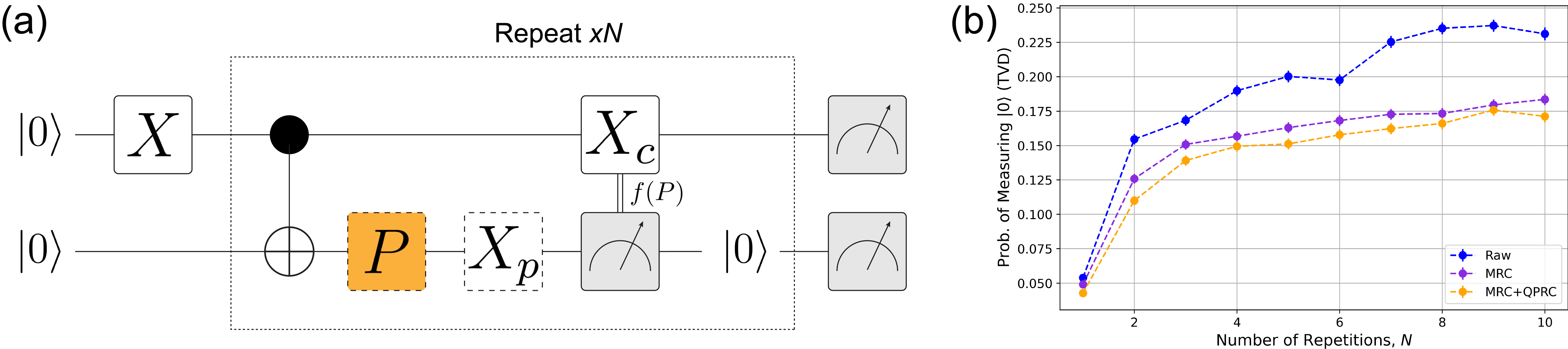}
    \caption{\textbf{Quasi-Probabilistic Readout Correction of Mid-Circuit Measurements.} 
    \textbf{(a)} Schematic for active bit-flip protection. A memory qubit [top] is prepared in the $\ket{1}$ state and a CNOT gate is applied between the data qubit and an ancilla qubit [bottom], which is subsequently measured. A conditional bit-flip ($X_c$) is performed on the memory qubit depending on the results of the MCM of the ancilla qubit, after which the ancilla qubit is reset. This process is performed for $N$ repetitions to protect the memory qubit from decaying to the ground state. Under MRC, the MCM is performed with RC by insertion of random Paulis ($P$, dashed orange box) before the MCM, in which case the conditional operation on the memory qubit is a function of the Pauli inserted before measurement, $f(P)$. To perform quasi-probabilistic readout correction on the MCM, a Pauli-$X$ gate is probabilistically inserted before the measurement ($X_p$, dashed white box), and the final results of the MCM with $X_p$ are subtracted from the results without $X_p$.
    \textbf{(b)} Results from performing the scheme presented in (a) for a 10 rounds of bit-flip protection. In the bare case, the probability of measuring the memory qubit in the $\ket{0}$ state grows from $\sim$5\% to $\sim$23\%. When the MCM is performed with MRC it only grows to $\sim$18\%. When the results measured with $X_p$ are subtracted from the results measured without $X_p$ (MRC+QPRC), the probability of measuring the memory qubit in $\ket{0}$ reduces by $\sim$1\% for all $N$ compared to MRC alone. This is consistent with a measured bit-flip rate of 1.1\% for the ancilla qubit. (Error bars are on the order of the size of the markers.)
    }
    \label{fig:fig4}
\end{figure*}

The QPRC protocol presented in the previous section provides a clear strategy for correcting readout errors afflicting terminating measurements. However, it is less clear how to correct readout errors in mid-circuit measurements (MCMs), which are subject to complex error processes that are not always present for terminating measurements \cite{rudinger2022characterizing}, and whose results can be used to adapt circuits in real-time via classical feedback \cite{foss2023experimental}. While the result of a single measurement used for decision branching in feed-forward schemes cannot be corrected in real-time, the results of MCMs can still be corrected quasi-probabilistically in the paradigm where we end up with an ensemble distribution at the very end of a circuit. To do so requires characterizing the probability of bit-flips for a given MCM, and quasi-probabilistically cancelling this error via random insertion of artificial Pauli-$X$ errors. We describe this procedure below.

When MCMs are used to perform conditional feed-forward operations, the readout fidelity of each MCM will dictate the rate at which incorrect decision branching occurs, and thus the rate at which the incorrect conditional operation is performed, which will add up linearly as a function of the number of MCMs in the circuit. In a model in which readout errors are purely probabilistic, this rate can be measured \textit{a priori} by characterizing the probability of a bit-flip error on the measured qubit(s). For example, suppose a qubit prepared in the ground state has a probability $p_1$ of experiencing a bit-flip during a MCM, then the probability with which a single instance of the MCM performs the correct conditional operation is $1 - p_1 = p_0$. According to the QPRC protocol presented in the previous section, the results of an imperfect measurement can be corrected by assigning a negative weight to the incorrect outcomes and subtracting them from the ideal outcomes. To do so in circuits with MCMs, we probabilistically insert an artificial bit-flip $X_p$ prior to the measured qubit with probability $p = p_1$. Now, for a circuit measured $N_s$ times, on average the correct conditional operation will have been applied $(1 - p) N_s$ times, and the incorrect conditional operation will have been applied $p N_s$ times. To mitigate the impact of the noisy MCM, we subtract the raw counts of the circuit measured with $X_p$ from the raw counts of the circuit measured without $X_p$. For circuits with multiple rounds of MCMs, we assign a negative weight to each instance in which $X_p$ appears in the circuit; thus, for odd (even) occurrences, the results are subtracted (added) to the bare results. This process generally increases the shot noise, since the error mitigated results only have $(1 - p) N_s - p N_s = (1 - 2p) N_s$ shots; one can choose to compensate for this at the cost of a larger overhead by increasing the total number of shots to $N_s' = N_s / (1 - 2p)$.

We demonstrate the correction of readout errors on MCMs by performing the above protocol on a circuit designed to protect the memory of a qubit in the $\ket{1}$ state, shown in \fig\ref{fig:fig4}a. Real-time active feedback is performed using the open-source control hardware \texttt{QubiC} \cite{xu2021qubic, xu2023qubic}. When MRC is utilized for MCMs, the conditional readout value of the measured qubit now depends on the Pauli that is sampled before readout, and thus the conditional operation on the memory qubit is a function of this Pauli, $f(P)$; the random sampling of $P$ and the calculation of $f(P)$ are performed many times before runtime, but the conditional bit-flip $X_c$ is performed in real-time with a feedback latency of 150 ns. In \fig\ref{fig:fig4}b, we plot the probability of measuring the memory qubit in $\ket{0}$ as a function of the number ($N$) of rounds of MCMs. We find that for the raw output, the probability is $\sim$5\% for $N=1$, growing to $\sim$23\% for $N=10$. When we perform the MCMs with MRC, this probability is reduced significantly, growing to only $\sim$18\% for $N=10$ rounds of MCMs. This difference can be explained by the fact that, in the ideal scenario, the ancilla qubit should be in $\ket{1}$ before each MCM --- indicating that the memory qubit has not experience a bit-flip --- which normally has a lower readout fidelity than the ground state. However, with MRC the readout fidelities are equal [$p(0|0) = p(1|1)$], so the error in the raw output will grow faster than the results with MRC. Additionally, inserting a random Pauli before the measurement can decouple the ancilla qubit from the memory qubit, ensuring that the error grows smoothly and monotonically. Finally, when we add QPRC to the MCMs performed with MRC, this reduces the probability of measuring the memory qubit in $\ket{0}$ by $\sim$1\% compared to just using MRC. The difference between MRC and MRC + QPRC is consistent with a bit-flip rate of 1.1\% measured for the ancilla qubit prior to the experiment. 

It should be noted that Ref.~\cite{gupta2023probabilistic} proposes a related method for mitigating Pauli errors that occur during MCMs using a quasi-probabilistic error cancellation scheme that depends on randomized compiling \cite{ferracin2022efficiently}. There are advantages and disadvantages to both techniques. A key distinction is that the protocol in \cite{gupta2023probabilistic} utilizes cycle benchmarking \cite{erhard2019characterizing} to characterize the rates of Pauli errors, which has a much higher characterization overhead than our technique --- whose characterization overhead is constant in the number of qubits $n$ --- and depends on the ability to measure exponential decays for the cycle or subcircuit of interest. However, the protocol in \cite{gupta2023probabilistic} can mitigate global errors that occur across an entire register of qubits, including correlated gate errors, whereas QPRC is only designed to mitigate readout errors. Future work could explore the scalability and trade-offs between these related methods as they relate to MCMs and adaptive circuits.

%% file: sections/5_conclusions.tex
\section{Discussion}\label{sec:discussion}

Improving the fidelity of qubit readout is equally as important as improving gates fidelities. However, in recent years, much more focus has been placed on improving gate fidelities, leaving readout errors (or more generally SPAM errors) much larger than contemporary gate errors. To compensate for this, experimentalists typical correct readout errors by inverting a $2^n \times 2^n$ confusion matrix or, alternatively, inverting local $2 \times 2$ readout confusion matrices for each qubit independently. While the former method can correct $n$-qubit readout errors that occur on computational basis states, it is not scalable; on the other hand, while the latter method is scalable, it cannot correct correlated readout errors. 

In this work, we introduce a quasi-probabilistic method for correcting measurement noise which utilizes randomized compiling for enforcing a stochastic bit-flip model of readout errors. Our method requires a minimal characterization overhead which is constant in the number of qubits, and is scalable in the limit that probabilistically-small readout errors can be ignored. We demonstrate that our method outperforms both full and local readout correction on a large number of different possible input states for eight qubits. Moreover, we show that it can be extended to scenarios where MCMs are used in the single-shot limit for adaptive feedback, as long as the end goal is to collect ensemble statistics of the outputs.

While significant research and development is required to improve the readout fidelities of contemporary qubits on many hardware platforms, scalable, matrix-inversion-free readout correction methods such as QPRC are useful tools for correcting readout errors in the NISQ era and beyond. Our method is fully compatible with MCMs, and future work could demonstrate the utility of utilizing QPRC for correcting readout errors in adaptive circuits used for preparing non-local entangled states \cite{hashim2024efficient}. Furthermore, the machinery needed for adaptive circuits is the same as what is needed for quantum error correction, so combining QPRC with quantum error correction would be an intriguing avenue for exploration.

\emph{Note added} --- During the completion of this manuscript, we became aware of a related but independently developed error-mitigation technique for mid-circuit measurements which appeared at the same time~\cite{ivashkov2023povm}.

%% file: sections/acknowledgements.tex
\section*{Acknowledgements} \label{sec:acknowledgements}
This work was supported by the U.S.~Department of Energy, Office of Science, Office of Advanced Scientific Computing Research Quantum Testbed Program under Contract No.~DE-AC02-05CH11231.

A.H.~and A.C.D.~designed the experiments and analyzed the data. A.C.D.~developed the quasi-probabilistic readout correction protocol. L.C.~and C.J.~fabricated the sample. N.F., Y.X., and G.H.~developed the classical control hardware used in this work. J.J.W~and I.S.~supervised all work.

A.H.~acknowledges fruitful discussions with Samuele Ferracin, Jan Balewski, Senrui Chen, and Liang Jiang.

%% file: sections/appendices.tex
\setcounter{table}{0}
\renewcommand{\thetable}{A\arabic{table}}

\setcounter{figure}{0}
\renewcommand{\thefigure}{A\arabic{figure}}


\begin{figure*}[th]
    \centering
    \includegraphics[width=2\columnwidth]{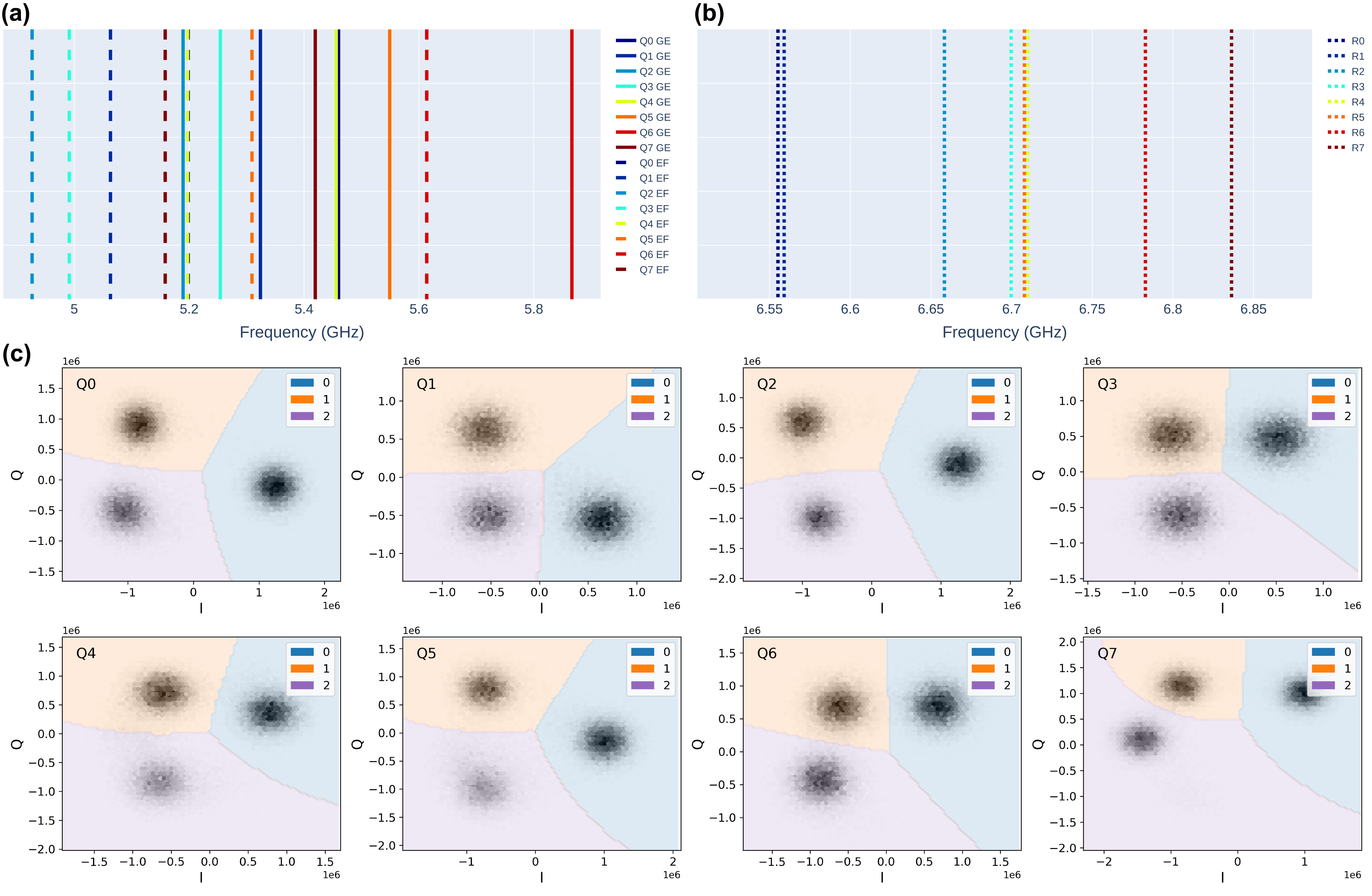}
    \caption{\textbf{Frequencies and Readout Characterization.}
        (a) Frequency spectrum of the GE (solid lines) and EF (dashed lines) transitions of the eight qubits on the quantum processor.
        (b) Frequency spectrum for the readout resonators coupled to the qubits.
        (c) Qutrit state discrimination is supported for all qubits on the quantum processor.
    }
    \label{fig:qubit_ro}
\end{figure*}

\input{tables/qubit_ro_gate_properties}

\section{Qubit \& Readout Characterization}\label{sec:appendix}

The quantum processing unit (QPU) used in this work consists of eight superconducting transmon qubits arranged in a ring geometry (\fig\ref{fig:fig1}a). The frequency spectrum of the GE and EF transition of each qubit is plotted in \fig\ref{fig:qubit_ro}a. Some frequency crowding is observed at the lower end of the frequency spectrum. For example, the GE transition of Q2 is close to the EF transitions of Q0, Q4, and Q3. This can lead to microwave line crosstalk between qubits, which can result in coherent leakage on the EF transitions when the GE transition of Q2 is driven. A similar effect can occur between the GE transition of Q5 and the EF transition of Q6, which is spectrally far from the rest of the qubits on the QPU due to fabrication inaccuracies. The qubit coherences are listed in Table \ref{tab:qubit_properties}.

In \fig\ref{fig:qubit_ro}c, we plot the readout calibration for all eight qubits on this QPU, which supports qutrit state discrimination. In qubit computations, qutrit readout can be used to measure leakage rates. Alternatively, qutrit state discrimination can be used for excited state promotion (ESP) \cite{mallet2009single, elder2020high} for improving qubit readout fidelities, whereby a $\pi_{1 \rightarrow 2}$ pulse is applied to each qubit before readout, after which all $\ket{2}$ state results are reclassified as $\ket{1}$ in post-processing. ESP can protect qubits against energy relaxation during readout, which can include readout-induced decay. We utilize ESP to improve qubit readout, and calibrate readout amplitudes to maximize readout fidelity with ESP turned on. The simultaneous readout fidelities are listed in Table \ref{tab:ro_fid_esp}.

Even with improved readout fidelities using ESP, qubits can experience readout crosstalk during measurement. In \fig\ref{fig:qubit_ro}b, we plot the frequency spectrum of the readout resonators for all eight qubits. We observe that several readout resonators are close in frequency. For example, the readout resonators for Q0 and Q1 are within $\sim$4 MHz of each other, and the readout resonators for Q3, Q4, and Q5 are all within $\sim$11 MHz of each other. Readout crosstalk can lead to context-dependent readout errors, in which the error on one qubit depends on the state of another qubit. This effect is apparent in the results presented in \fig\ref{fig:fig1}, in which the $\ket{011}$ and $\ket{111}$ states have drastically worse readout fidelities than the other preparation states.

\section{Gate Benchmarking}\label{sec:gate_benchmarking}

The single-qubit gates and two-qubit gates used in this worked are benchmarked using randomized benchmarking (RB) and cycle benchmarking (CB). Infidelities for single-qubit gates are listed in Table \ref{tab:single_qubit_gate_inf}. Infidelities for two-qubit gates are listed in Table \ref{tab:two_qubit_gate_inf}. It should be noted that all quoted infidelities are the \emph{process infidelity} $e_F$, not the \emph{average gate infidelity} $r$. These two are equal up to a simple dimensionality factor:
\begin{equation}
    e_F = \frac{d+1}{d} r \;,
\end{equation}
where $d = 2^n$ for $n$ qubits.

\section{Quantum Hardware}\label{sec:hardware}

In this work, we use the open-source control system \texttt{QubiC} \cite{xu2021qubic, xu2023qubic} to perform these experiments. \texttt{QubiC} is an FPGA-based control system for superconducting qubits developed at Lawrence Berkeley National Lab. The \texttt{QubiC} system used for these experiments was implemented on the Xilinx ZCU216 RFSoC (RF system-on-chip) evaluation board, and uses custom gateware for real-time pulse sequencing and synthesis. 

The \texttt{QubiC} gateware has a bank of distributed processor cores for performing pulse sequencing, parameterization, and conditional execution (i.e., control flow) \cite{fruitwala2024distributed}. The \texttt{QubiC} readout DSP (digital signal processing) chain includes on-FPGA demodulation and qubit state discrimination using a threshold mechanism. Currently, the discrimination is performed for MCMs using the in-phase ($I$) component of the integrated readout pulse. If $I > 0$, the discriminator returns a 0; if $I < 0$, the discriminator returns a 1. For this reason, all of the $\ket{0}$ states are calibrated to be on the right side of $I = 0$, and all of the $\ket{1}$ and $\ket{2}$ states are calibrated to be on the left side of $I = 0$ (see \fig\ref{fig:qubit_ro}c). These state-discriminated results can then be requested by any processor core (using a special instruction) and used as inputs to arbitrary control flow/branching decisions (e.g., a \texttt{while} loop or \texttt{if/else} code block). The total feedback latency (not including readout time) is 150 ns. After these experiments were performed, a neural network-based readout discriminator was developed for MCMs performed on \texttt{QubiC} which is capable of distinguishing $\ket{1}$ from $\ket{2}$ \cite{vora2024ml}.

\section{Scaling the Quasi-Probabilistic Inverse For a Higher Number of Qubits}\label{sec:generalized_inversion}

In Sec.~\ref{sec:qp}, we propose to invert the effect of measurement errors by applying an approximate inverse
operation described in \eq\ref{eq:qp_dist_k}. While this worked well in our experiment, such an inversion technique may
fail when the error probability grows above $1/2$. Without a workaround, this constraint would make it impossible for 
such a quasi-probabilistic inversion technique to scale, since the error probability grows exponentially in the number of qubits.
Indeed, with independent local error rates of $2\%$, the total error probability reaches $1/2$ after 34 qubits. As such, 
in this section, we provide a generalization of our quasi-probabilistic inversion technique that ensures
a proper scaling of our mitigation method for measurements.

To explain the generalization, let us consider a two-qubit toy example where the error probability distribution takes the form:
\begin{align}
    p_{AB}(\epsilon) =& \left(\frac{9}{16}-\epsilon\right) \bs{00} + \frac{3}{16} \bs{01}+ \frac{3}{16} \bs{10} \notag \\&+ \left(\frac{1}{16}+\epsilon\right) \bs{11} ~,
\end{align}
where $\epsilon \in [0,1/4)$, and where the underscore notation indicates a classical bit-string in the probability distribution. Notice that in the case where $\epsilon=0$, we fall back on marginalized independent bit-flip errors of $25\%$. As such, $\epsilon$ denotes a correlated bit-flip error probability on top of the independent bit-flips. Notice that for all values of $\epsilon > 1/16$, the total error probability exceeds $1/2$. It should be noted that the first-order inversion makes things worse when 
\begin{align}
    \abs{\frac{p_{\0}^2}{2p_{\0}-1}-1}<\abs{p_{\0}-1} ~,
\end{align}
which occurs when $p_{\0} < 2/3$.

As we demonstrate here, even in the milder case where $\epsilon = 0$, $p_{\0} < 2/3$, and
the first-order inversion described in \cref{eq:qp_dist},
\begin{align}
    q_{AB}(\epsilon = 0) = \frac{9}{2} \bs{00} - \frac{3}{2} \bs{01}- \frac{3}{2} \bs{10} - \frac{1}{2} \bs{11}~,
\end{align}
fails to mitigate errors:
\begin{align}
    q_{AB}( 0 ) \oplus p_{AB}(0 ) = \frac{31}{16} \bs{00} - \frac{3}{16} \bs{01} - \frac{3}{16} \bs{10} - \frac{9}{16} \bs{11}~.
\end{align}

However ---  and this is the essence of the solution --- it is possible to adapt the quasi-probabilistic inversion 
to address the qubits $A$ and $B$ individually before mitigating them in tandem. Let us consider the error distributions marginalized over individual qubits:
\begin{subequations}
\begin{align}
    p_A(\epsilon) &= \left(\frac{3}{4} -\epsilon\right) \bsmarg{0}{A} +  \left(\frac{1}{4} + \epsilon\right) \bsmarg{1}{A}~, \\
    p_B(\epsilon) &= \left(\frac{3}{4} -\epsilon\right) \bsmarg{0}{B} +  \left(\frac{1}{4} + \epsilon\right) \bsmarg{1}{B}~.
\end{align}
\end{subequations}
Notice that the error probabilities marginalized on the two subsystems are below $1/3$ as long as $\epsilon$ remains below $1/12$. From there, consider the first-order local quasi-probabilistic inversions obtained by using \eq\ref{eq:qp_dist} with the marginalized distributions $p_A(\epsilon)$ and $p_B(\epsilon)$:
\begin{subequations}
\begin{align}
    q_A(\epsilon) &= \frac{2}{1-4\epsilon} \left[\left(\frac{3}{4} -\epsilon\right) \bsmarg{0}{A} -  \left(\frac{1}{4} + \epsilon\right) \bsmarg{1}{A} \right]~, \\
    q_B(\epsilon) &= \frac{2}{1-4\epsilon} \left[\left(\frac{3}{4} -\epsilon\right) \bsmarg{0}{B} -  \left(\frac{1}{4} + \epsilon\right) \bsmarg{1}{B} \right]~.
\end{align}
\end{subequations}
Let us apply those local inversions to the total distribution $p_{AB}(\epsilon)$. Unsurprisingly, in the case where $\epsilon = 0$, the error distribution is locally independent [i.e., $p_{AB}(0) = p_A(0) p_B(0)$], and
we obtain a perfect error mitigation:
\begin{align}
    q_A(0)q_B(0) \oplus p_{AB}(0) = \bs{00}~.
\end{align}
In the more general case, we get:
\begin{align}
    &p'_{AB}(\epsilon) = q_A(\epsilon)q_B(\epsilon) \oplus p_{AB}(\epsilon) = \left( \frac{3 + (1-4\epsilon)^{-2}}{4}\right) \bs{00} \notag \\
    &+ \left( \frac{1- (1-4\epsilon)^{-2}}{4}\right) \bs{01} + \left( \frac{1- (1-4\epsilon)^{-2}}{4}\right) \bs{10} \notag \\
    &+ \left( \frac{-1+(1-4\epsilon)^{-2}}{4}\right) \bs{11}~.
\end{align}
Instead of further carrying out heavy expressions in $\epsilon$, for the sake of simplicity (especially since this a toy example), let us pick $\epsilon$ small enough such that we can ignore $\mathcal{O}(\epsilon^2)$. In that case, we get
\begin{align}
    p'_{AB}(\epsilon)  = (1+2\epsilon) \bs{00} -2\epsilon \bs{01} -2\epsilon \bs{10}+ 2\epsilon \bs{11} + \mathcal{O}(\epsilon^2)~.
\end{align}
Let us now apply the quasi-probabilistic inversion technique prescribed in Sec.~\ref{sec:qp} on that new (quasi-probabilistic) distribution. Using
\begin{align}
    q'_{AB}(\epsilon)  = \frac{1}{1+4\epsilon}\Big( (1+2\epsilon) \bs{00} +2\epsilon \bs{01} &+2\epsilon \bs{10}\notag \\
    &- 2\epsilon \bs{11} \Big)~,
\end{align}
we get
\begin{align}
    q'_{AB}(\epsilon) \oplus p'_{AB}(\epsilon)   = \bs{00} +O(\epsilon^2)~.
\end{align}
In other words, performing the local mitigation $q_A(\epsilon)q_B(\epsilon)$ changed the error distribution and brought us to a point where we could effectively apply a joint mitigation operation.

Notice that in this toy example, we applied the quasi-probabilistic inversions sequentially,
but they can easily be compiled into a single operation:
\begin{align}
    &q_{AB}(\epsilon):=q'_{AB}(\epsilon) \oplus q_A(\epsilon)q_B(\epsilon)   \notag \\
    &= \left(\frac{9}{4}+4 \epsilon\right)\bs{00} -\frac{3}{4}\bs{01} -\frac{3}{4}\bs{10} +\left( \frac{1}{4}-4 \epsilon\right)\bs{11}+O(\epsilon^2)~.
\end{align}
This is relevant to readout correction of mid-circuit measurements, where the quasi-probability $q_{AB}(\epsilon)$ 
is used as input for sampling circuits (see section Sec.~\ref{sec:mcm}). 

Under the light of the toy example, the generalization of the mitigation strategy for larger systems is fairly straightforward. Given an error distribution $p_S$ over a set of qubits $S$, subdivide the system into disjoint partitions $S_1, \cdots, S_k$ such that $\bigcup_i S_i = S$ and obtain the marginal error probabilities over those partitions, $p_{S_1}, \cdots, p_{S_k}$. The goal is to choose partitioning such that the marginal error probability within every partition is lower than $1/2$ (or $1/3$ if using first-order inversions). From these marginal error distributions, apply quasi-probabilistic corrections according to the method described in Sec.~\ref{sec:qp}:
\begin{align}
    p_S \rightarrow  q_{S_1} \cdots q_{S_k} \oplus p_S  = p'_S~.
\end{align}
$p'_S$ should be closer to the identity. Repeat the process for increasingly larger 
partitions. Once the resulting total error quasi-distribution is close enough to the identity
(e.g., once $|1-p_{\0}|<1/2$), use the total distribution to infer the 
quasi-probabilistic inverse.

\section{Measurement RC and Quasi-Probabilistic Readout Correction on Eight Qubits}\label{sec:8q_mrc_qprc}

\begin{figure*}[th]
    \centering
    \includegraphics[width=2\columnwidth]{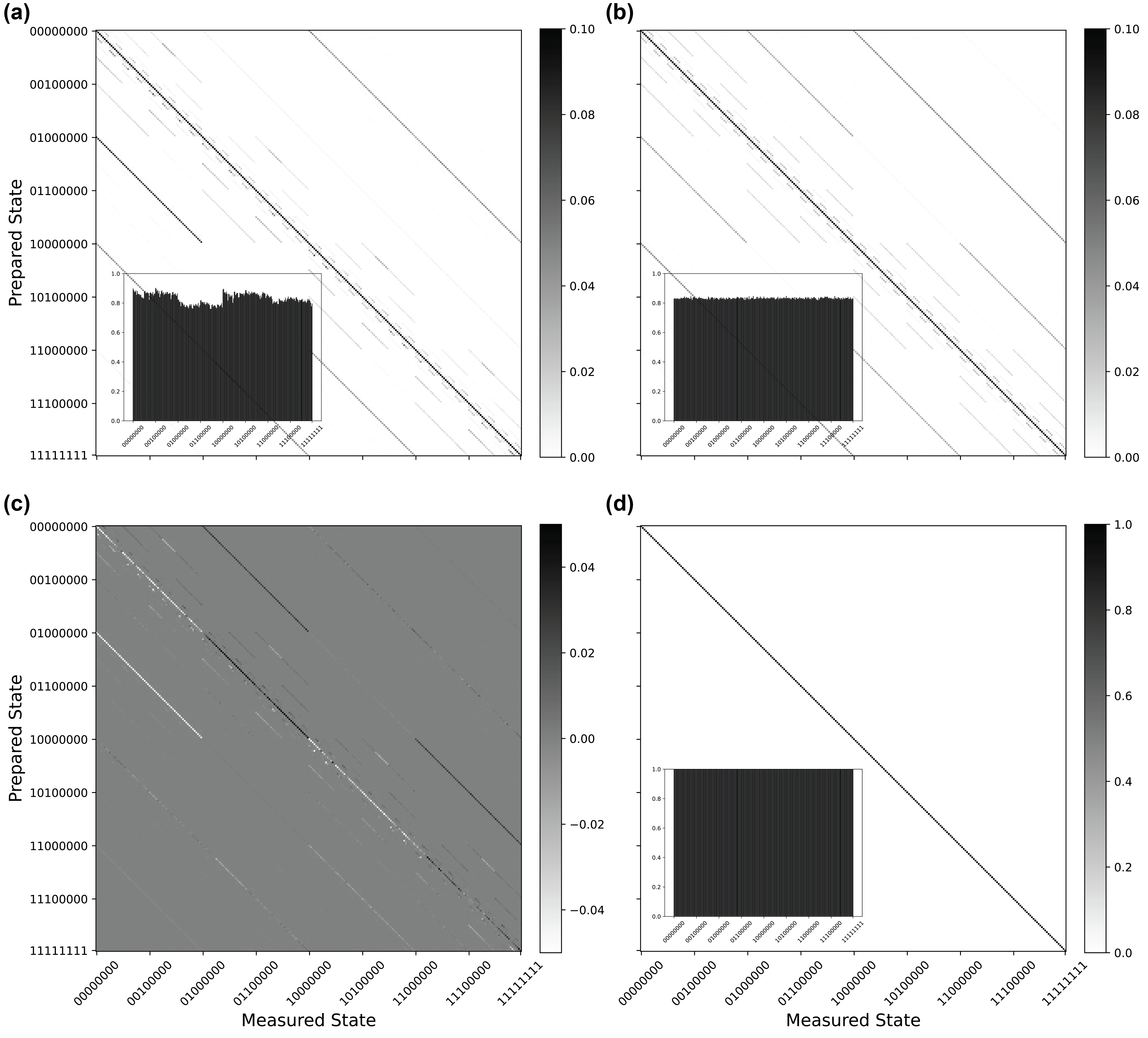}
    \caption{\textbf{MRC and QPRC on Eight Qubits.}
    (a) Eight-qubit confusion matrix. The maximum value of the colormap is set to 0.1 for better contrast of the off-diagonal elements. Inset: diagonal values of the confusion matrix (i.e., readout fidelities for different states).
    (b) Eight-qubit confusion matrix measured with MRC. The maximum value of the colormap is set to 0.1 for better contrast of the off-diagonal elements. Inset: diagonal values of the confusion matrix (i.e., readout fidelities for different states).
    (c) The data from (b) minus the data from (a). This difference shows which entries are increased (black) or decreased (white) by using MRC.
    (d) Eight-qubit confusion matrix reconstruction from (b) by applying QPRC to the measured data for each different state preparation. Inset: diagonal values of the confusion matrix (i.e., effective readout fidelities for different states). The perfect correction for all states demonstrates that QPRC can correct readout errors equally well regardless of which input state was initially characterized.
    }
    \label{fig:8q_qpmrc}
\end{figure*}

In the main body of the paper, we show the impact of MRC on a confusion matrix for only three qubits (see \fig\ref{fig:fig1}). To demonstrate that MRC scales equally well to larger numbers of qubits, we apply MRC to a full eight-qubit confusion matrix, shown in \fig\ref{fig:8q_qpmrc}. In \fig\ref{fig:8q_qpmrc}(a), we plot the raw eight-qubit confusion matrix reconstructed for all qubits on our quantum processor. We observe an asymmetry in the diagonal and off-diagonal elements, indicating the presence of state and context-dependent errors. In \fig\ref{fig:8q_qpmrc}(b), we plot the eight-qubit confusion matrix reconstructed with MRC. We observe symmetry in both the diagonal and off-diagonal elements, demonstrating that MRC works equally well in tailoring readout noise for eight qubits as it did for three qubits. In \fig\ref{fig:8q_qpmrc}(c), we plot the difference between the confusion matrix with MRC and the raw confusion matrix, showing where elements of the confusion matrix are increased or decreased with MRC. Finally, in \fig\ref{fig:8q_qpmrc}(d), we perform QPRC on the data in (b), to demonstrate that our effective readout fidelity is perfect across all states with QPRC. More specifically, for each preparation state, we apply QPRC to the measured data using the measured data itself to compute the inverse distribution, similar to the analysis performed in \fig\ref{fig:fig3}(b). This demonstrates that we can perform the QPRC protocol using readout errors characterized for any initial input state.

%% file: tables/qubit_ro_gate_properties.tex
\begin{table}[!ht]
\renewcommand{\arraystretch}{1.5}

\centering
\resizebox{\columnwidth}{!}{
\begin{tabular}{l | r r r r r r r r }
    \hline
    \hline
    {} & Q0 & Q1 & Q2 & Q3 & Q4 & Q5 & Q6 & Q7 \\
    \hline
    $T_1$ ($\mu$s) & 96.6(2.6) & 130.0(2.7) & 142.0(3.0) & 140.0(6.3) & 77.0(5.2) & 30.4(0.95) & 55.6(1.3) & 22.5(0.32) \\
    $T_{2}^*$ ($\mu$s) & 120.0(14.0) & 41.0(7.2) & 92.0(16.0) & 61.0(6.1) & 38.0(5.4) & 8.5(1.3) & 26.0(3.7) & 39.0(1.7) \\
    $T_{2E}$ ($\mu$s) & 120.0(8.3) & 130.0(7.5) & 140.0(12.0) & 90.0(13.0) & 110.0(11.0) & 33.0(3.6) & 90.0(14.0) & 43.0(2.2) \\
    \hline
    \hline
\end{tabular}}
\caption{\textbf{Qubit Coherences.} Qubit coherence times ($T_1$, $T_{2}^*$, $T_{2E}$) are listed above.}
\label{tab:qubit_properties}

\bigskip

\centering
\resizebox{\columnwidth}{!}{
\begin{tabular}{l | r r r r r r r r }
    \hline
    \hline
    {} & Q0 & Q1 & Q2 & Q3 & Q4 & Q5 & Q6 & Q7 \\
    \hline
    $P(0 | 0)$ & 0.995(1) & 0.995(1) & 0.995(1) & 0.992(2) & 0.990(2) & 0.998(1) & 0.997(1) & 0.987(3) \\
    $P(1 | 1)$ & 0.983(2) & 0.962(7) & 0.994(2) & 0.986(2) & 0.966(7) & 0.969(4) & 0.994(2) & 0.986(2) \\
    \hline
    \hline
\end{tabular}}
\caption{\textbf{Readout Fidelities.}
Simultaneous readout fidelities for all qubits with excited state promotion.}
\label{tab:ro_fid_esp}

\bigskip

\centering
\resizebox{\columnwidth}{!}{
\begin{tabular}{l | r r r r r r r r }
    \hline
    \hline
    {} & Q0 & Q1 & Q2 & Q3 & Q4 & Q5 & Q6 & Q7 \\
    \hline
    RB iso. ($10^{-3}$) & 1.5(1) & 0.33(2) & 0.54(3) & 1.0(1) & 3.2(1) & 1.92(9) & 2.4(3) & 2.58(9) \\
    RB sim. ($10^{-3}$) & 2.1(2) & 3.1(2) & 2.0(3) & 1.9(2) & 6.1(7) & 5.7(3) & 3.4(3) & 7.6(9) \\
    \hline
    \hline
\end{tabular}}
\caption{\textbf{Single-qubit Gate Infidelities.}
    The process infidelities for isolated and simultaneous single-qubit gates measured via RB for each qubit are listed above. 
    }
\label{tab:single_qubit_gate_inf}

\bigskip

\centering
\resizebox{\columnwidth}{!}{
\begin{tabular}{l | r r r r r r r r }
    \hline
    \hline
    {} & (Q0, Q1) & (Q1, Q2) & (Q2, Q3) & (Q3, Q4) & (Q4, Q5) & (Q5, Q6) & (Q6, Q7) & (Q7, Q0) \\
    \hline
    RB iso. ($10^{-2}$) & 5.0(5) & 1.62(7) & 1.64(8) & 2.6(2) & 3.8(2) & 4.6(2) & 6.7(6) & 3.8(3) \\
    CB (CZ) ($10^{-2}$) & 1.4(1) & 0.57(1) & 0.41(1) & 0.81(4) & 1.80(8) & 2.08(3) & 2.77(8) & 1.40(7)  \\
    \hline
    \hline
\end{tabular}}
\caption{\textbf{Two-qubit Gate Infidelities.}
    The process infidelities for two-qubit RB are listed above for each qubit pair used in this work. Native (CZ) gate fidelities are measured via CB.
}
\label{tab:two_qubit_gate_inf}



\end{table}

%% file: main.bbl
\begin{thebibliography}{49}%
\makeatletter
\providecommand \@ifxundefined [1]{%
 \@ifx{#1\undefined}
}%
\providecommand \@ifnum [1]{%
 \ifnum #1\expandafter \@firstoftwo
 \else \expandafter \@secondoftwo
 \fi
}%
\providecommand \@ifx [1]{%
 \ifx #1\expandafter \@firstoftwo
 \else \expandafter \@secondoftwo
 \fi
}%
\providecommand \natexlab [1]{#1}%
\providecommand \enquote  [1]{``#1''}%
\providecommand \bibnamefont  [1]{#1}%
\providecommand \bibfnamefont [1]{#1}%
\providecommand \citenamefont [1]{#1}%
\providecommand \href@noop [0]{\@secondoftwo}%
\providecommand \href [0]{\begingroup \@sanitize@url \@href}%
\providecommand \@href[1]{\@@startlink{#1}\@@href}%
\providecommand \@@href[1]{\endgroup#1\@@endlink}%
\providecommand \@sanitize@url [0]{\catcode `\\12\catcode `\$12\catcode `\&12\catcode `\#12\catcode `\^12\catcode `\_12\catcode `\%12\relax}%
\providecommand \@@startlink[1]{}%
\providecommand \@@endlink[0]{}%
\providecommand \url  [0]{\begingroup\@sanitize@url \@url }%
\providecommand \@url [1]{\endgroup\@href {#1}{\urlprefix }}%
\providecommand \urlprefix  [0]{URL }%
\providecommand \Eprint [0]{\href }%
\providecommand \doibase [0]{http://dx.doi.org/}%
\providecommand \selectlanguage [0]{\@gobble}%
\providecommand \bibinfo  [0]{\@secondoftwo}%
\providecommand \bibfield  [0]{\@secondoftwo}%
\providecommand \translation [1]{[#1]}%
\providecommand \BibitemOpen [0]{}%
\providecommand \bibitemStop [0]{}%
\providecommand \bibitemNoStop [0]{.\EOS\space}%
\providecommand \EOS [0]{\spacefactor3000\relax}%
\providecommand \BibitemShut  [1]{\csname bibitem#1\endcsname}%
\let\auto@bib@innerbib\@empty
\bibitem [{\citenamefont {Gottesman}\ and\ \citenamefont {Chuang}(1999)}]{gottesman1999demonstrating}%
  \BibitemOpen
  \bibfield  {author} {\bibinfo {author} {\bibfnamefont {D.}~\bibnamefont {Gottesman}}\ and\ \bibinfo {author} {\bibfnamefont {I.~L.}\ \bibnamefont {Chuang}},\ }\href@noop {} {\bibfield  {journal} {\bibinfo  {journal} {Nature}\ }\textbf {\bibinfo {volume} {402}},\ \bibinfo {pages} {390} (\bibinfo {year} {1999})}\BibitemShut {NoStop}%
\bibitem [{\citenamefont {Chou}\ \emph {et~al.}(2018)\citenamefont {Chou}, \citenamefont {Blumoff}, \citenamefont {Wang}, \citenamefont {Reinhold}, \citenamefont {Axline}, \citenamefont {Gao}, \citenamefont {Frunzio}, \citenamefont {Devoret}, \citenamefont {Jiang},\ and\ \citenamefont {Schoelkopf}}]{chou2018deterministic}%
  \BibitemOpen
  \bibfield  {author} {\bibinfo {author} {\bibfnamefont {K.~S.}\ \bibnamefont {Chou}}, \bibinfo {author} {\bibfnamefont {J.~Z.}\ \bibnamefont {Blumoff}}, \bibinfo {author} {\bibfnamefont {C.~S.}\ \bibnamefont {Wang}}, \bibinfo {author} {\bibfnamefont {P.~C.}\ \bibnamefont {Reinhold}}, \bibinfo {author} {\bibfnamefont {C.~J.}\ \bibnamefont {Axline}}, \bibinfo {author} {\bibfnamefont {Y.~Y.}\ \bibnamefont {Gao}}, \bibinfo {author} {\bibfnamefont {L.}~\bibnamefont {Frunzio}}, \bibinfo {author} {\bibfnamefont {M.}~\bibnamefont {Devoret}}, \bibinfo {author} {\bibfnamefont {L.}~\bibnamefont {Jiang}}, \ and\ \bibinfo {author} {\bibfnamefont {R.}~\bibnamefont {Schoelkopf}},\ }\href@noop {} {\bibfield  {journal} {\bibinfo  {journal} {Nature}\ }\textbf {\bibinfo {volume} {561}},\ \bibinfo {pages} {368} (\bibinfo {year} {2018})}\BibitemShut {NoStop}%
\bibitem [{\citenamefont {Raussendorf}\ and\ \citenamefont {Briegel}(2001)}]{raussendorf2001one}%
  \BibitemOpen
  \bibfield  {author} {\bibinfo {author} {\bibfnamefont {R.}~\bibnamefont {Raussendorf}}\ and\ \bibinfo {author} {\bibfnamefont {H.~J.}\ \bibnamefont {Briegel}},\ }\href@noop {} {\bibfield  {journal} {\bibinfo  {journal} {Physical review letters}\ }\textbf {\bibinfo {volume} {86}},\ \bibinfo {pages} {5188} (\bibinfo {year} {2001})}\BibitemShut {NoStop}%
\bibitem [{\citenamefont {Briegel}\ \emph {et~al.}(2009)\citenamefont {Briegel}, \citenamefont {Browne}, \citenamefont {D{\"u}r}, \citenamefont {Raussendorf},\ and\ \citenamefont {Van~den Nest}}]{briegel2009measurement}%
  \BibitemOpen
  \bibfield  {author} {\bibinfo {author} {\bibfnamefont {H.~J.}\ \bibnamefont {Briegel}}, \bibinfo {author} {\bibfnamefont {D.~E.}\ \bibnamefont {Browne}}, \bibinfo {author} {\bibfnamefont {W.}~\bibnamefont {D{\"u}r}}, \bibinfo {author} {\bibfnamefont {R.}~\bibnamefont {Raussendorf}}, \ and\ \bibinfo {author} {\bibfnamefont {M.}~\bibnamefont {Van~den Nest}},\ }\href@noop {} {\bibfield  {journal} {\bibinfo  {journal} {Nature Physics}\ }\textbf {\bibinfo {volume} {5}},\ \bibinfo {pages} {19} (\bibinfo {year} {2009})}\BibitemShut {NoStop}%
\bibitem [{\citenamefont {Foss-Feig}\ \emph {et~al.}(2023)\citenamefont {Foss-Feig}, \citenamefont {Tikku}, \citenamefont {Lu}, \citenamefont {Mayer}, \citenamefont {Iqbal}, \citenamefont {Gatterman}, \citenamefont {Gerber}, \citenamefont {Gilmore}, \citenamefont {Gresh}, \citenamefont {Hankin} \emph {et~al.}}]{foss2023experimental}%
  \BibitemOpen
  \bibfield  {author} {\bibinfo {author} {\bibfnamefont {M.}~\bibnamefont {Foss-Feig}}, \bibinfo {author} {\bibfnamefont {A.}~\bibnamefont {Tikku}}, \bibinfo {author} {\bibfnamefont {T.-C.}\ \bibnamefont {Lu}}, \bibinfo {author} {\bibfnamefont {K.}~\bibnamefont {Mayer}}, \bibinfo {author} {\bibfnamefont {M.}~\bibnamefont {Iqbal}}, \bibinfo {author} {\bibfnamefont {T.~M.}\ \bibnamefont {Gatterman}}, \bibinfo {author} {\bibfnamefont {J.~A.}\ \bibnamefont {Gerber}}, \bibinfo {author} {\bibfnamefont {K.}~\bibnamefont {Gilmore}}, \bibinfo {author} {\bibfnamefont {D.}~\bibnamefont {Gresh}}, \bibinfo {author} {\bibfnamefont {A.}~\bibnamefont {Hankin}},  \emph {et~al.},\ }\href@noop {} {\bibfield  {journal} {\bibinfo  {journal} {arXiv preprint arXiv:2302.03029}\ } (\bibinfo {year} {2023})}\BibitemShut {NoStop}%
\bibitem [{\citenamefont {Hashim}\ \emph {et~al.}(2024)\citenamefont {Hashim}, \citenamefont {Yuan}, \citenamefont {Gokhale}, \citenamefont {Chen}, \citenamefont {Juenger}, \citenamefont {Fruitwala}, \citenamefont {Xu}, \citenamefont {Huang}, \citenamefont {Jiang},\ and\ \citenamefont {Siddiqi}}]{hashim2024efficient}%
  \BibitemOpen
  \bibfield  {author} {\bibinfo {author} {\bibfnamefont {A.}~\bibnamefont {Hashim}}, \bibinfo {author} {\bibfnamefont {M.}~\bibnamefont {Yuan}}, \bibinfo {author} {\bibfnamefont {P.}~\bibnamefont {Gokhale}}, \bibinfo {author} {\bibfnamefont {L.}~\bibnamefont {Chen}}, \bibinfo {author} {\bibfnamefont {C.}~\bibnamefont {Juenger}}, \bibinfo {author} {\bibfnamefont {N.}~\bibnamefont {Fruitwala}}, \bibinfo {author} {\bibfnamefont {Y.}~\bibnamefont {Xu}}, \bibinfo {author} {\bibfnamefont {G.}~\bibnamefont {Huang}}, \bibinfo {author} {\bibfnamefont {L.}~\bibnamefont {Jiang}}, \ and\ \bibinfo {author} {\bibfnamefont {I.}~\bibnamefont {Siddiqi}},\ }\href@noop {} {\bibfield  {journal} {\bibinfo  {journal} {arXiv preprint arXiv:2403.18768}\ } (\bibinfo {year} {2024})}\BibitemShut {NoStop}%
\bibitem [{\citenamefont {Shor}(1995)}]{shor1995scheme}%
  \BibitemOpen
  \bibfield  {author} {\bibinfo {author} {\bibfnamefont {P.~W.}\ \bibnamefont {Shor}},\ }\href@noop {} {\bibfield  {journal} {\bibinfo  {journal} {Physical review A}\ }\textbf {\bibinfo {volume} {52}},\ \bibinfo {pages} {R2493} (\bibinfo {year} {1995})}\BibitemShut {NoStop}%
\bibitem [{\citenamefont {Knill}\ and\ \citenamefont {Laflamme}(1997)}]{knill1997theory}%
  \BibitemOpen
  \bibfield  {author} {\bibinfo {author} {\bibfnamefont {E.}~\bibnamefont {Knill}}\ and\ \bibinfo {author} {\bibfnamefont {R.}~\bibnamefont {Laflamme}},\ }\href@noop {} {\bibfield  {journal} {\bibinfo  {journal} {Physical Review A}\ }\textbf {\bibinfo {volume} {55}},\ \bibinfo {pages} {900} (\bibinfo {year} {1997})}\BibitemShut {NoStop}%
\bibitem [{\citenamefont {Chiaverini}\ \emph {et~al.}(2004)\citenamefont {Chiaverini}, \citenamefont {Leibfried}, \citenamefont {Schaetz}, \citenamefont {Barrett}, \citenamefont {Blakestad}, \citenamefont {Britton}, \citenamefont {Itano}, \citenamefont {Jost}, \citenamefont {Knill}, \citenamefont {Langer} \emph {et~al.}}]{chiaverini2004realization}%
  \BibitemOpen
  \bibfield  {author} {\bibinfo {author} {\bibfnamefont {J.}~\bibnamefont {Chiaverini}}, \bibinfo {author} {\bibfnamefont {D.}~\bibnamefont {Leibfried}}, \bibinfo {author} {\bibfnamefont {T.}~\bibnamefont {Schaetz}}, \bibinfo {author} {\bibfnamefont {M.~D.}\ \bibnamefont {Barrett}}, \bibinfo {author} {\bibfnamefont {R.}~\bibnamefont {Blakestad}}, \bibinfo {author} {\bibfnamefont {J.}~\bibnamefont {Britton}}, \bibinfo {author} {\bibfnamefont {W.~M.}\ \bibnamefont {Itano}}, \bibinfo {author} {\bibfnamefont {J.~D.}\ \bibnamefont {Jost}}, \bibinfo {author} {\bibfnamefont {E.}~\bibnamefont {Knill}}, \bibinfo {author} {\bibfnamefont {C.}~\bibnamefont {Langer}},  \emph {et~al.},\ }\href@noop {} {\bibfield  {journal} {\bibinfo  {journal} {Nature}\ }\textbf {\bibinfo {volume} {432}},\ \bibinfo {pages} {602} (\bibinfo {year} {2004})}\BibitemShut {NoStop}%
\bibitem [{\citenamefont {Ofek}\ \emph {et~al.}(2016)\citenamefont {Ofek}, \citenamefont {Petrenko}, \citenamefont {Heeres}, \citenamefont {Reinhold}, \citenamefont {Leghtas}, \citenamefont {Vlastakis}, \citenamefont {Liu}, \citenamefont {Frunzio}, \citenamefont {Girvin}, \citenamefont {Jiang} \emph {et~al.}}]{ofek2016extending}%
  \BibitemOpen
  \bibfield  {author} {\bibinfo {author} {\bibfnamefont {N.}~\bibnamefont {Ofek}}, \bibinfo {author} {\bibfnamefont {A.}~\bibnamefont {Petrenko}}, \bibinfo {author} {\bibfnamefont {R.}~\bibnamefont {Heeres}}, \bibinfo {author} {\bibfnamefont {P.}~\bibnamefont {Reinhold}}, \bibinfo {author} {\bibfnamefont {Z.}~\bibnamefont {Leghtas}}, \bibinfo {author} {\bibfnamefont {B.}~\bibnamefont {Vlastakis}}, \bibinfo {author} {\bibfnamefont {Y.}~\bibnamefont {Liu}}, \bibinfo {author} {\bibfnamefont {L.}~\bibnamefont {Frunzio}}, \bibinfo {author} {\bibfnamefont {S.}~\bibnamefont {Girvin}}, \bibinfo {author} {\bibfnamefont {L.}~\bibnamefont {Jiang}},  \emph {et~al.},\ }\href@noop {} {\bibfield  {journal} {\bibinfo  {journal} {Nature}\ }\textbf {\bibinfo {volume} {536}},\ \bibinfo {pages} {441} (\bibinfo {year} {2016})}\BibitemShut {NoStop}%
\bibitem [{\citenamefont {Rosenblum}\ \emph {et~al.}(2018)\citenamefont {Rosenblum}, \citenamefont {Reinhold}, \citenamefont {Mirrahimi}, \citenamefont {Jiang}, \citenamefont {Frunzio},\ and\ \citenamefont {Schoelkopf}}]{rosenblum2018fault}%
  \BibitemOpen
  \bibfield  {author} {\bibinfo {author} {\bibfnamefont {S.}~\bibnamefont {Rosenblum}}, \bibinfo {author} {\bibfnamefont {P.}~\bibnamefont {Reinhold}}, \bibinfo {author} {\bibfnamefont {M.}~\bibnamefont {Mirrahimi}}, \bibinfo {author} {\bibfnamefont {L.}~\bibnamefont {Jiang}}, \bibinfo {author} {\bibfnamefont {L.}~\bibnamefont {Frunzio}}, \ and\ \bibinfo {author} {\bibfnamefont {R.~J.}\ \bibnamefont {Schoelkopf}},\ }\href@noop {} {\bibfield  {journal} {\bibinfo  {journal} {Science}\ }\textbf {\bibinfo {volume} {361}},\ \bibinfo {pages} {266} (\bibinfo {year} {2018})}\BibitemShut {NoStop}%
\bibitem [{\citenamefont {Livingston}\ \emph {et~al.}(2022)\citenamefont {Livingston}, \citenamefont {Blok}, \citenamefont {Flurin}, \citenamefont {Dressel}, \citenamefont {Jordan},\ and\ \citenamefont {Siddiqi}}]{livingston2022experimental}%
  \BibitemOpen
  \bibfield  {author} {\bibinfo {author} {\bibfnamefont {W.~P.}\ \bibnamefont {Livingston}}, \bibinfo {author} {\bibfnamefont {M.~S.}\ \bibnamefont {Blok}}, \bibinfo {author} {\bibfnamefont {E.}~\bibnamefont {Flurin}}, \bibinfo {author} {\bibfnamefont {J.}~\bibnamefont {Dressel}}, \bibinfo {author} {\bibfnamefont {A.~N.}\ \bibnamefont {Jordan}}, \ and\ \bibinfo {author} {\bibfnamefont {I.}~\bibnamefont {Siddiqi}},\ }\href@noop {} {\bibfield  {journal} {\bibinfo  {journal} {Nature communications}\ }\textbf {\bibinfo {volume} {13}},\ \bibinfo {pages} {2307} (\bibinfo {year} {2022})}\BibitemShut {NoStop}%
\bibitem [{\citenamefont {Blumoff}\ \emph {et~al.}(2016)\citenamefont {Blumoff}, \citenamefont {Chou}, \citenamefont {Shen}, \citenamefont {Reagor}, \citenamefont {Axline}, \citenamefont {Brierley}, \citenamefont {Silveri}, \citenamefont {Wang}, \citenamefont {Vlastakis}, \citenamefont {Nigg} \emph {et~al.}}]{blumoff2016implementing}%
  \BibitemOpen
  \bibfield  {author} {\bibinfo {author} {\bibfnamefont {J.~Z.}\ \bibnamefont {Blumoff}}, \bibinfo {author} {\bibfnamefont {K.}~\bibnamefont {Chou}}, \bibinfo {author} {\bibfnamefont {C.}~\bibnamefont {Shen}}, \bibinfo {author} {\bibfnamefont {M.}~\bibnamefont {Reagor}}, \bibinfo {author} {\bibfnamefont {C.}~\bibnamefont {Axline}}, \bibinfo {author} {\bibfnamefont {R.}~\bibnamefont {Brierley}}, \bibinfo {author} {\bibfnamefont {M.}~\bibnamefont {Silveri}}, \bibinfo {author} {\bibfnamefont {C.}~\bibnamefont {Wang}}, \bibinfo {author} {\bibfnamefont {B.}~\bibnamefont {Vlastakis}}, \bibinfo {author} {\bibfnamefont {S.~E.}\ \bibnamefont {Nigg}},  \emph {et~al.},\ }\href@noop {} {\bibfield  {journal} {\bibinfo  {journal} {Physical Review X}\ }\textbf {\bibinfo {volume} {6}},\ \bibinfo {pages} {031041} (\bibinfo {year} {2016})}\BibitemShut {NoStop}%
\bibitem [{\citenamefont {Heinsoo}\ \emph {et~al.}(2018)\citenamefont {Heinsoo}, \citenamefont {Andersen}, \citenamefont {Remm}, \citenamefont {Krinner}, \citenamefont {Walter}, \citenamefont {Salath{\'e}}, \citenamefont {Gasparinetti}, \citenamefont {Besse}, \citenamefont {Poto{\v{c}}nik}, \citenamefont {Wallraff} \emph {et~al.}}]{heinsoo2018rapid}%
  \BibitemOpen
  \bibfield  {author} {\bibinfo {author} {\bibfnamefont {J.}~\bibnamefont {Heinsoo}}, \bibinfo {author} {\bibfnamefont {C.~K.}\ \bibnamefont {Andersen}}, \bibinfo {author} {\bibfnamefont {A.}~\bibnamefont {Remm}}, \bibinfo {author} {\bibfnamefont {S.}~\bibnamefont {Krinner}}, \bibinfo {author} {\bibfnamefont {T.}~\bibnamefont {Walter}}, \bibinfo {author} {\bibfnamefont {Y.}~\bibnamefont {Salath{\'e}}}, \bibinfo {author} {\bibfnamefont {S.}~\bibnamefont {Gasparinetti}}, \bibinfo {author} {\bibfnamefont {J.-C.}\ \bibnamefont {Besse}}, \bibinfo {author} {\bibfnamefont {A.}~\bibnamefont {Poto{\v{c}}nik}}, \bibinfo {author} {\bibfnamefont {A.}~\bibnamefont {Wallraff}},  \emph {et~al.},\ }\href@noop {} {\bibfield  {journal} {\bibinfo  {journal} {Physical Review Applied}\ }\textbf {\bibinfo {volume} {10}},\ \bibinfo {pages} {034040} (\bibinfo {year} {2018})}\BibitemShut {NoStop}%
\bibitem [{\citenamefont {Chen}\ \emph {et~al.}(2019)\citenamefont {Chen}, \citenamefont {Farahzad}, \citenamefont {Yoo},\ and\ \citenamefont {Wei}}]{chen2019detector}%
  \BibitemOpen
  \bibfield  {author} {\bibinfo {author} {\bibfnamefont {Y.}~\bibnamefont {Chen}}, \bibinfo {author} {\bibfnamefont {M.}~\bibnamefont {Farahzad}}, \bibinfo {author} {\bibfnamefont {S.}~\bibnamefont {Yoo}}, \ and\ \bibinfo {author} {\bibfnamefont {T.-C.}\ \bibnamefont {Wei}},\ }\href@noop {} {\bibfield  {journal} {\bibinfo  {journal} {Physical Review A}\ }\textbf {\bibinfo {volume} {100}},\ \bibinfo {pages} {052315} (\bibinfo {year} {2019})}\BibitemShut {NoStop}%
\bibitem [{\citenamefont {Maciejewski}\ \emph {et~al.}(2020)\citenamefont {Maciejewski}, \citenamefont {Zimbor{\'a}s},\ and\ \citenamefont {Oszmaniec}}]{maciejewski2020mitigation}%
  \BibitemOpen
  \bibfield  {author} {\bibinfo {author} {\bibfnamefont {F.~B.}\ \bibnamefont {Maciejewski}}, \bibinfo {author} {\bibfnamefont {Z.}~\bibnamefont {Zimbor{\'a}s}}, \ and\ \bibinfo {author} {\bibfnamefont {M.}~\bibnamefont {Oszmaniec}},\ }\href@noop {} {\bibfield  {journal} {\bibinfo  {journal} {Quantum}\ }\textbf {\bibinfo {volume} {4}},\ \bibinfo {pages} {257} (\bibinfo {year} {2020})}\BibitemShut {NoStop}%
\bibitem [{\citenamefont {Elder}\ \emph {et~al.}(2020)\citenamefont {Elder}, \citenamefont {Wang}, \citenamefont {Reinhold}, \citenamefont {Hann}, \citenamefont {Chou}, \citenamefont {Lester}, \citenamefont {Rosenblum}, \citenamefont {Frunzio}, \citenamefont {Jiang},\ and\ \citenamefont {Schoelkopf}}]{elder2020high}%
  \BibitemOpen
  \bibfield  {author} {\bibinfo {author} {\bibfnamefont {S.~S.}\ \bibnamefont {Elder}}, \bibinfo {author} {\bibfnamefont {C.~S.}\ \bibnamefont {Wang}}, \bibinfo {author} {\bibfnamefont {P.}~\bibnamefont {Reinhold}}, \bibinfo {author} {\bibfnamefont {C.~T.}\ \bibnamefont {Hann}}, \bibinfo {author} {\bibfnamefont {K.~S.}\ \bibnamefont {Chou}}, \bibinfo {author} {\bibfnamefont {B.~J.}\ \bibnamefont {Lester}}, \bibinfo {author} {\bibfnamefont {S.}~\bibnamefont {Rosenblum}}, \bibinfo {author} {\bibfnamefont {L.}~\bibnamefont {Frunzio}}, \bibinfo {author} {\bibfnamefont {L.}~\bibnamefont {Jiang}}, \ and\ \bibinfo {author} {\bibfnamefont {R.~J.}\ \bibnamefont {Schoelkopf}},\ }\href@noop {} {\bibfield  {journal} {\bibinfo  {journal} {Physical Review X}\ }\textbf {\bibinfo {volume} {10}},\ \bibinfo {pages} {011001} (\bibinfo {year} {2020})}\BibitemShut {NoStop}%
\bibitem [{\citenamefont {Graydon}\ \emph {et~al.}(2022)\citenamefont {Graydon}, \citenamefont {Skanes-Norman},\ and\ \citenamefont {Wallman}}]{graydon2022designing}%
  \BibitemOpen
  \bibfield  {author} {\bibinfo {author} {\bibfnamefont {M.~A.}\ \bibnamefont {Graydon}}, \bibinfo {author} {\bibfnamefont {J.}~\bibnamefont {Skanes-Norman}}, \ and\ \bibinfo {author} {\bibfnamefont {J.~J.}\ \bibnamefont {Wallman}},\ }\href@noop {} {\bibfield  {journal} {\bibinfo  {journal} {arXiv preprint arXiv:2201.07156}\ } (\bibinfo {year} {2022})}\BibitemShut {NoStop}%
\bibitem [{\citenamefont {Wallman}\ and\ \citenamefont {Emerson}(2016)}]{wallman2016noise}%
  \BibitemOpen
  \bibfield  {author} {\bibinfo {author} {\bibfnamefont {J.~J.}\ \bibnamefont {Wallman}}\ and\ \bibinfo {author} {\bibfnamefont {J.}~\bibnamefont {Emerson}},\ }\href {\doibase 10.1103/PhysRevA.94.052325} {\bibfield  {journal} {\bibinfo  {journal} {Phys. Rev. A}\ }\textbf {\bibinfo {volume} {94}},\ \bibinfo {pages} {052325} (\bibinfo {year} {2016})}\BibitemShut {NoStop}%
\bibitem [{\citenamefont {Hashim}\ \emph {et~al.}(2021)\citenamefont {Hashim}, \citenamefont {Naik}, \citenamefont {Morvan}, \citenamefont {Ville}, \citenamefont {Mitchell}, \citenamefont {Kreikebaum}, \citenamefont {Davis}, \citenamefont {Smith}, \citenamefont {Iancu}, \citenamefont {O'Brien}, \citenamefont {Hincks}, \citenamefont {Wallman}, \citenamefont {Emerson},\ and\ \citenamefont {Siddiqi}}]{hashim2021randomized}%
  \BibitemOpen
  \bibfield  {author} {\bibinfo {author} {\bibfnamefont {A.}~\bibnamefont {Hashim}}, \bibinfo {author} {\bibfnamefont {R.~K.}\ \bibnamefont {Naik}}, \bibinfo {author} {\bibfnamefont {A.}~\bibnamefont {Morvan}}, \bibinfo {author} {\bibfnamefont {J.-L.}\ \bibnamefont {Ville}}, \bibinfo {author} {\bibfnamefont {B.}~\bibnamefont {Mitchell}}, \bibinfo {author} {\bibfnamefont {J.~M.}\ \bibnamefont {Kreikebaum}}, \bibinfo {author} {\bibfnamefont {M.}~\bibnamefont {Davis}}, \bibinfo {author} {\bibfnamefont {E.}~\bibnamefont {Smith}}, \bibinfo {author} {\bibfnamefont {C.}~\bibnamefont {Iancu}}, \bibinfo {author} {\bibfnamefont {K.~P.}\ \bibnamefont {O'Brien}}, \bibinfo {author} {\bibfnamefont {I.}~\bibnamefont {Hincks}}, \bibinfo {author} {\bibfnamefont {J.~J.}\ \bibnamefont {Wallman}}, \bibinfo {author} {\bibfnamefont {J.}~\bibnamefont {Emerson}}, \ and\ \bibinfo {author} {\bibfnamefont {I.}~\bibnamefont {Siddiqi}},\ }\href {\doibase 10.1103/PhysRevX.11.041039} {\bibfield  {journal} {\bibinfo  {journal} {Phys. Rev.
  X}\ }\textbf {\bibinfo {volume} {11}},\ \bibinfo {pages} {041039} (\bibinfo {year} {2021})}\BibitemShut {NoStop}%
\bibitem [{\citenamefont {Beale}\ and\ \citenamefont {Wallman}(2023)}]{beale2023randomized}%
  \BibitemOpen
  \bibfield  {author} {\bibinfo {author} {\bibfnamefont {S.~J.}\ \bibnamefont {Beale}}\ and\ \bibinfo {author} {\bibfnamefont {J.~J.}\ \bibnamefont {Wallman}},\ }\href@noop {} {\bibfield  {journal} {\bibinfo  {journal} {arXiv preprint arXiv:2304.06599}\ } (\bibinfo {year} {2023})}\BibitemShut {NoStop}%
\bibitem [{\citenamefont {Hashim}\ \emph {et~al.}(2023)\citenamefont {Hashim}, \citenamefont {Seritan}, \citenamefont {Proctor}, \citenamefont {Rudinger}, \citenamefont {Goss}, \citenamefont {Naik}, \citenamefont {Kreikebaum}, \citenamefont {Santiago},\ and\ \citenamefont {Siddiqi}}]{hashim2023benchmarking}%
  \BibitemOpen
  \bibfield  {author} {\bibinfo {author} {\bibfnamefont {A.}~\bibnamefont {Hashim}}, \bibinfo {author} {\bibfnamefont {S.}~\bibnamefont {Seritan}}, \bibinfo {author} {\bibfnamefont {T.}~\bibnamefont {Proctor}}, \bibinfo {author} {\bibfnamefont {K.}~\bibnamefont {Rudinger}}, \bibinfo {author} {\bibfnamefont {N.}~\bibnamefont {Goss}}, \bibinfo {author} {\bibfnamefont {R.}~\bibnamefont {Naik}}, \bibinfo {author} {\bibfnamefont {J.~M.}\ \bibnamefont {Kreikebaum}}, \bibinfo {author} {\bibfnamefont {D.}~\bibnamefont {Santiago}}, \ and\ \bibinfo {author} {\bibfnamefont {I.}~\bibnamefont {Siddiqi}},\ }\href {\doibase 10.1038/s41534-023-00764-y} {\bibfield  {journal} {\bibinfo  {journal} {npj Quantum Inf}\ }\textbf {\bibinfo {volume} {9}} (\bibinfo {year} {2023}),\ 10.1038/s41534-023-00764-y}\BibitemShut {NoStop}%
\bibitem [{\citenamefont {McLaren}\ \emph {et~al.}(2023)\citenamefont {McLaren}, \citenamefont {Graydon},\ and\ \citenamefont {Wallman}}]{mclaren2023stochastic}%
  \BibitemOpen
  \bibfield  {author} {\bibinfo {author} {\bibfnamefont {D.}~\bibnamefont {McLaren}}, \bibinfo {author} {\bibfnamefont {M.~A.}\ \bibnamefont {Graydon}}, \ and\ \bibinfo {author} {\bibfnamefont {J.~J.}\ \bibnamefont {Wallman}},\ }\href@noop {} {\bibfield  {journal} {\bibinfo  {journal} {arXiv preprint arXiv:2306.07418}\ } (\bibinfo {year} {2023})}\BibitemShut {NoStop}%
\bibitem [{\citenamefont {Bravyi}\ \emph {et~al.}(2021)\citenamefont {Bravyi}, \citenamefont {Sheldon}, \citenamefont {Kandala}, \citenamefont {Mckay},\ and\ \citenamefont {Gambetta}}]{bravyi2021mitigating}%
  \BibitemOpen
  \bibfield  {author} {\bibinfo {author} {\bibfnamefont {S.}~\bibnamefont {Bravyi}}, \bibinfo {author} {\bibfnamefont {S.}~\bibnamefont {Sheldon}}, \bibinfo {author} {\bibfnamefont {A.}~\bibnamefont {Kandala}}, \bibinfo {author} {\bibfnamefont {D.~C.}\ \bibnamefont {Mckay}}, \ and\ \bibinfo {author} {\bibfnamefont {J.~M.}\ \bibnamefont {Gambetta}},\ }\href@noop {} {\bibfield  {journal} {\bibinfo  {journal} {Physical Review A}\ }\textbf {\bibinfo {volume} {103}},\ \bibinfo {pages} {042605} (\bibinfo {year} {2021})}\BibitemShut {NoStop}%
\bibitem [{\citenamefont {G{\"u}nther}\ \emph {et~al.}(2021)\citenamefont {G{\"u}nther}, \citenamefont {Tacchino}, \citenamefont {Wootton}, \citenamefont {Tavernelli},\ and\ \citenamefont {Barkoutsos}}]{gunther2021improving}%
  \BibitemOpen
  \bibfield  {author} {\bibinfo {author} {\bibfnamefont {J.~M.}\ \bibnamefont {G{\"u}nther}}, \bibinfo {author} {\bibfnamefont {F.}~\bibnamefont {Tacchino}}, \bibinfo {author} {\bibfnamefont {J.~R.}\ \bibnamefont {Wootton}}, \bibinfo {author} {\bibfnamefont {I.}~\bibnamefont {Tavernelli}}, \ and\ \bibinfo {author} {\bibfnamefont {P.~K.}\ \bibnamefont {Barkoutsos}},\ }\href@noop {} {\bibfield  {journal} {\bibinfo  {journal} {Quantum Science and Technology}\ }\textbf {\bibinfo {volume} {7}},\ \bibinfo {pages} {015009} (\bibinfo {year} {2021})}\BibitemShut {NoStop}%
\bibitem [{\citenamefont {Hicks}\ \emph {et~al.}(2022)\citenamefont {Hicks}, \citenamefont {Kobrin}, \citenamefont {Bauer},\ and\ \citenamefont {Nachman}}]{hicks2022active}%
  \BibitemOpen
  \bibfield  {author} {\bibinfo {author} {\bibfnamefont {R.}~\bibnamefont {Hicks}}, \bibinfo {author} {\bibfnamefont {B.}~\bibnamefont {Kobrin}}, \bibinfo {author} {\bibfnamefont {C.~W.}\ \bibnamefont {Bauer}}, \ and\ \bibinfo {author} {\bibfnamefont {B.}~\bibnamefont {Nachman}},\ }\href@noop {} {\bibfield  {journal} {\bibinfo  {journal} {Physical Review A}\ }\textbf {\bibinfo {volume} {105}},\ \bibinfo {pages} {012419} (\bibinfo {year} {2022})}\BibitemShut {NoStop}%
\bibitem [{\citenamefont {{Wallman}}(2018)}]{Wallman2018}%
  \BibitemOpen
  \bibfield  {author} {\bibinfo {author} {\bibfnamefont {J.~J.}\ \bibnamefont {{Wallman}}},\ }\href {\doibase 10.22331/q-2018-01-29-47} {\bibfield  {journal} {\bibinfo  {journal} {Quantum}\ }\textbf {\bibinfo {volume} {2}},\ \bibinfo {pages} {47} (\bibinfo {year} {2018})},\ \Eprint {http://arxiv.org/abs/1703.09835} {arXiv:1703.09835 [quant-ph]} \BibitemShut {NoStop}%
\bibitem [{\citenamefont {{Winick}}\ \emph {et~al.}(2022)\citenamefont {{Winick}}, \citenamefont {{Wallman}}, \citenamefont {{Dahlen}}, \citenamefont {{Hincks}}, \citenamefont {{Ospadov}},\ and\ \citenamefont {{Emerson}}}]{Winick2022}%
  \BibitemOpen
  \bibfield  {author} {\bibinfo {author} {\bibfnamefont {A.}~\bibnamefont {{Winick}}}, \bibinfo {author} {\bibfnamefont {J.~J.}\ \bibnamefont {{Wallman}}}, \bibinfo {author} {\bibfnamefont {D.}~\bibnamefont {{Dahlen}}}, \bibinfo {author} {\bibfnamefont {I.}~\bibnamefont {{Hincks}}}, \bibinfo {author} {\bibfnamefont {E.}~\bibnamefont {{Ospadov}}}, \ and\ \bibinfo {author} {\bibfnamefont {J.}~\bibnamefont {{Emerson}}},\ }\href {\doibase 10.48550/arXiv.2212.07500} {\bibfield  {journal} {\bibinfo  {journal} {arXiv e-prints}\ ,\ \bibinfo {eid} {arXiv:2212.07500}} (\bibinfo {year} {2022})},\ \Eprint {http://arxiv.org/abs/2212.07500} {arXiv:2212.07500 [quant-ph]} \BibitemShut {NoStop}%
\bibitem [{\citenamefont {Goss}\ \emph {et~al.}(2023)\citenamefont {Goss}, \citenamefont {Ferracin}, \citenamefont {Hashim}, \citenamefont {Carignan-Dugas}, \citenamefont {Kreikebaum}, \citenamefont {Naik}, \citenamefont {Santiago},\ and\ \citenamefont {Siddiqi}}]{goss2023extending}%
  \BibitemOpen
  \bibfield  {author} {\bibinfo {author} {\bibfnamefont {N.}~\bibnamefont {Goss}}, \bibinfo {author} {\bibfnamefont {S.}~\bibnamefont {Ferracin}}, \bibinfo {author} {\bibfnamefont {A.}~\bibnamefont {Hashim}}, \bibinfo {author} {\bibfnamefont {A.}~\bibnamefont {Carignan-Dugas}}, \bibinfo {author} {\bibfnamefont {J.~M.}\ \bibnamefont {Kreikebaum}}, \bibinfo {author} {\bibfnamefont {R.~K.}\ \bibnamefont {Naik}}, \bibinfo {author} {\bibfnamefont {D.~I.}\ \bibnamefont {Santiago}}, \ and\ \bibinfo {author} {\bibfnamefont {I.}~\bibnamefont {Siddiqi}},\ }\href@noop {} {\bibfield  {journal} {\bibinfo  {journal} {arXiv preprint arXiv:2305.16507}\ } (\bibinfo {year} {2023})}\BibitemShut {NoStop}%
\bibitem [{\citenamefont {Smith}\ \emph {et~al.}(2021)\citenamefont {Smith}, \citenamefont {Khosla}, \citenamefont {Self},\ and\ \citenamefont {Kim}}]{smith2021qubit}%
  \BibitemOpen
  \bibfield  {author} {\bibinfo {author} {\bibfnamefont {A.~W.}\ \bibnamefont {Smith}}, \bibinfo {author} {\bibfnamefont {K.~E.}\ \bibnamefont {Khosla}}, \bibinfo {author} {\bibfnamefont {C.~N.}\ \bibnamefont {Self}}, \ and\ \bibinfo {author} {\bibfnamefont {M.}~\bibnamefont {Kim}},\ }\href@noop {} {\bibfield  {journal} {\bibinfo  {journal} {Science advances}\ }\textbf {\bibinfo {volume} {7}},\ \bibinfo {pages} {eabi8009} (\bibinfo {year} {2021})}\BibitemShut {NoStop}%
\bibitem [{\citenamefont {Hicks}\ \emph {et~al.}(2021)\citenamefont {Hicks}, \citenamefont {Bauer},\ and\ \citenamefont {Nachman}}]{hicks2021readout}%
  \BibitemOpen
  \bibfield  {author} {\bibinfo {author} {\bibfnamefont {R.}~\bibnamefont {Hicks}}, \bibinfo {author} {\bibfnamefont {C.~W.}\ \bibnamefont {Bauer}}, \ and\ \bibinfo {author} {\bibfnamefont {B.}~\bibnamefont {Nachman}},\ }\href@noop {} {\bibfield  {journal} {\bibinfo  {journal} {Physical Review A}\ }\textbf {\bibinfo {volume} {103}},\ \bibinfo {pages} {022407} (\bibinfo {year} {2021})}\BibitemShut {NoStop}%
\bibitem [{Note1()}]{Note1}%
  \BibitemOpen
  \bibinfo {note} {While Pauli-$Z$ gates can be implemented entirely via phase shifts in the following pulse, this will have no effect if the gate directly precedes a measurement. Instead, we implement Pauli-$Z$ gates using the $ZXZXZ$ decomposition of the gate, such that physical pulses are always played prior to measurement.}\BibitemShut {Stop}%
\bibitem [{\citenamefont {Van Den~Berg}\ \emph {et~al.}(2022)\citenamefont {Van Den~Berg}, \citenamefont {Minev},\ and\ \citenamefont {Temme}}]{van2022model}%
  \BibitemOpen
  \bibfield  {author} {\bibinfo {author} {\bibfnamefont {E.}~\bibnamefont {Van Den~Berg}}, \bibinfo {author} {\bibfnamefont {Z.~K.}\ \bibnamefont {Minev}}, \ and\ \bibinfo {author} {\bibfnamefont {K.}~\bibnamefont {Temme}},\ }\href@noop {} {\bibfield  {journal} {\bibinfo  {journal} {Physical Review A}\ }\textbf {\bibinfo {volume} {105}},\ \bibinfo {pages} {032620} (\bibinfo {year} {2022})}\BibitemShut {NoStop}%
\bibitem [{Note2()}]{Note2}%
  \BibitemOpen
  \bibinfo {note} {The accuracy of the estimate is also limited by state preparation errors and single-qubit gate errors, but these tend to be low compared to measurement errors.}\BibitemShut {Stop}%
\bibitem [{\citenamefont {Li}\ and\ \citenamefont {Benjamin}(2017)}]{ying2017efficient}%
  \BibitemOpen
  \bibfield  {author} {\bibinfo {author} {\bibfnamefont {Y.}~\bibnamefont {Li}}\ and\ \bibinfo {author} {\bibfnamefont {S.~C.}\ \bibnamefont {Benjamin}},\ }\href {\doibase 10.1103/PhysRevX.7.021050} {\bibfield  {journal} {\bibinfo  {journal} {Phys. Rev. X}\ }\textbf {\bibinfo {volume} {7}},\ \bibinfo {pages} {021050} (\bibinfo {year} {2017})}\BibitemShut {NoStop}%
\bibitem [{\citenamefont {Temme}\ \emph {et~al.}(2017)\citenamefont {Temme}, \citenamefont {Bravyi},\ and\ \citenamefont {Gambetta}}]{Temme2017mitigation}%
  \BibitemOpen
  \bibfield  {author} {\bibinfo {author} {\bibfnamefont {K.}~\bibnamefont {Temme}}, \bibinfo {author} {\bibfnamefont {S.}~\bibnamefont {Bravyi}}, \ and\ \bibinfo {author} {\bibfnamefont {J.~M.}\ \bibnamefont {Gambetta}},\ }\href {\doibase 10.1103/PhysRevLett.119.180509} {\bibfield  {journal} {\bibinfo  {journal} {Phys. Rev. Lett.}\ }\textbf {\bibinfo {volume} {119}},\ \bibinfo {pages} {180509} (\bibinfo {year} {2017})}\BibitemShut {NoStop}%
\bibitem [{\citenamefont {Endo}\ \emph {et~al.}(2018)\citenamefont {Endo}, \citenamefont {Benjamin},\ and\ \citenamefont {Li}}]{endo2018mitigation}%
  \BibitemOpen
  \bibfield  {author} {\bibinfo {author} {\bibfnamefont {S.}~\bibnamefont {Endo}}, \bibinfo {author} {\bibfnamefont {S.~C.}\ \bibnamefont {Benjamin}}, \ and\ \bibinfo {author} {\bibfnamefont {Y.}~\bibnamefont {Li}},\ }\href {\doibase 10.1103/PhysRevX.8.031027} {\bibfield  {journal} {\bibinfo  {journal} {Phys. Rev. X}\ }\textbf {\bibinfo {volume} {8}},\ \bibinfo {pages} {031027} (\bibinfo {year} {2018})}\BibitemShut {NoStop}%
\bibitem [{\citenamefont {Kandala}\ \emph {et~al.}(2019)\citenamefont {Kandala}, \citenamefont {Temme}, \citenamefont {Córcoles}, \citenamefont {Mezzacapo}, \citenamefont {Chow},\ and\ \citenamefont {Gambetta}}]{Kandala_2019}%
  \BibitemOpen
  \bibfield  {author} {\bibinfo {author} {\bibfnamefont {A.}~\bibnamefont {Kandala}}, \bibinfo {author} {\bibfnamefont {K.}~\bibnamefont {Temme}}, \bibinfo {author} {\bibfnamefont {A.~D.}\ \bibnamefont {Córcoles}}, \bibinfo {author} {\bibfnamefont {A.}~\bibnamefont {Mezzacapo}}, \bibinfo {author} {\bibfnamefont {J.~M.}\ \bibnamefont {Chow}}, \ and\ \bibinfo {author} {\bibfnamefont {J.~M.}\ \bibnamefont {Gambetta}},\ }\href {\doibase 10.1038/s41586-019-1040-7} {\bibfield  {journal} {\bibinfo  {journal} {Nature}\ }\textbf {\bibinfo {volume} {567}},\ \bibinfo {pages} {491–495} (\bibinfo {year} {2019})}\BibitemShut {NoStop}%
\bibitem [{\citenamefont {Ferracin}\ \emph {et~al.}(2022)\citenamefont {Ferracin}, \citenamefont {Hashim}, \citenamefont {Ville}, \citenamefont {Naik}, \citenamefont {Carignan-Dugas}, \citenamefont {Qassim}, \citenamefont {Morvan}, \citenamefont {Santiago}, \citenamefont {Siddiqi},\ and\ \citenamefont {Wallman}}]{ferracin2022efficiently}%
  \BibitemOpen
  \bibfield  {author} {\bibinfo {author} {\bibfnamefont {S.}~\bibnamefont {Ferracin}}, \bibinfo {author} {\bibfnamefont {A.}~\bibnamefont {Hashim}}, \bibinfo {author} {\bibfnamefont {J.-L.}\ \bibnamefont {Ville}}, \bibinfo {author} {\bibfnamefont {R.}~\bibnamefont {Naik}}, \bibinfo {author} {\bibfnamefont {A.}~\bibnamefont {Carignan-Dugas}}, \bibinfo {author} {\bibfnamefont {H.}~\bibnamefont {Qassim}}, \bibinfo {author} {\bibfnamefont {A.}~\bibnamefont {Morvan}}, \bibinfo {author} {\bibfnamefont {D.~I.}\ \bibnamefont {Santiago}}, \bibinfo {author} {\bibfnamefont {I.}~\bibnamefont {Siddiqi}}, \ and\ \bibinfo {author} {\bibfnamefont {J.~J.}\ \bibnamefont {Wallman}},\ }\href@noop {} {\bibfield  {journal} {\bibinfo  {journal} {arXiv preprint arXiv:2201.10672}\ } (\bibinfo {year} {2022})}\BibitemShut {NoStop}%
\bibitem [{\citenamefont {Takagi}\ \emph {et~al.}(2022)\citenamefont {Takagi}, \citenamefont {Endo}, \citenamefont {Minagawa},\ and\ \citenamefont {Gu}}]{Takagi_2022}%
  \BibitemOpen
  \bibfield  {author} {\bibinfo {author} {\bibfnamefont {R.}~\bibnamefont {Takagi}}, \bibinfo {author} {\bibfnamefont {S.}~\bibnamefont {Endo}}, \bibinfo {author} {\bibfnamefont {S.}~\bibnamefont {Minagawa}}, \ and\ \bibinfo {author} {\bibfnamefont {M.}~\bibnamefont {Gu}},\ }\href {\doibase 10.1038/s41534-022-00618-z} {\bibfield  {journal} {\bibinfo  {journal} {npj Quantum Information}\ }\textbf {\bibinfo {volume} {8}} (\bibinfo {year} {2022}),\ 10.1038/s41534-022-00618-z}\BibitemShut {NoStop}%
\bibitem [{\citenamefont {Rudinger}\ \emph {et~al.}(2022)\citenamefont {Rudinger}, \citenamefont {Ribeill}, \citenamefont {Govia}, \citenamefont {Ware}, \citenamefont {Nielsen}, \citenamefont {Young}, \citenamefont {Ohki}, \citenamefont {Blume-Kohout},\ and\ \citenamefont {Proctor}}]{rudinger2022characterizing}%
  \BibitemOpen
  \bibfield  {author} {\bibinfo {author} {\bibfnamefont {K.}~\bibnamefont {Rudinger}}, \bibinfo {author} {\bibfnamefont {G.~J.}\ \bibnamefont {Ribeill}}, \bibinfo {author} {\bibfnamefont {L.~C.}\ \bibnamefont {Govia}}, \bibinfo {author} {\bibfnamefont {M.}~\bibnamefont {Ware}}, \bibinfo {author} {\bibfnamefont {E.}~\bibnamefont {Nielsen}}, \bibinfo {author} {\bibfnamefont {K.}~\bibnamefont {Young}}, \bibinfo {author} {\bibfnamefont {T.~A.}\ \bibnamefont {Ohki}}, \bibinfo {author} {\bibfnamefont {R.}~\bibnamefont {Blume-Kohout}}, \ and\ \bibinfo {author} {\bibfnamefont {T.}~\bibnamefont {Proctor}},\ }\href@noop {} {\bibfield  {journal} {\bibinfo  {journal} {Physical Review Applied}\ }\textbf {\bibinfo {volume} {17}},\ \bibinfo {pages} {014014} (\bibinfo {year} {2022})}\BibitemShut {NoStop}%
\bibitem [{\citenamefont {Xu}\ \emph {et~al.}(2021)\citenamefont {Xu}, \citenamefont {Huang}, \citenamefont {Balewski}, \citenamefont {Naik}, \citenamefont {Morvan}, \citenamefont {Mitchell}, \citenamefont {Nowrouzi}, \citenamefont {Santiago},\ and\ \citenamefont {Siddiqi}}]{xu2021qubic}%
  \BibitemOpen
  \bibfield  {author} {\bibinfo {author} {\bibfnamefont {Y.}~\bibnamefont {Xu}}, \bibinfo {author} {\bibfnamefont {G.}~\bibnamefont {Huang}}, \bibinfo {author} {\bibfnamefont {J.}~\bibnamefont {Balewski}}, \bibinfo {author} {\bibfnamefont {R.}~\bibnamefont {Naik}}, \bibinfo {author} {\bibfnamefont {A.}~\bibnamefont {Morvan}}, \bibinfo {author} {\bibfnamefont {B.}~\bibnamefont {Mitchell}}, \bibinfo {author} {\bibfnamefont {K.}~\bibnamefont {Nowrouzi}}, \bibinfo {author} {\bibfnamefont {D.~I.}\ \bibnamefont {Santiago}}, \ and\ \bibinfo {author} {\bibfnamefont {I.}~\bibnamefont {Siddiqi}},\ }\href@noop {} {\bibfield  {journal} {\bibinfo  {journal} {IEEE Transactions on Quantum Engineering}\ }\textbf {\bibinfo {volume} {2}},\ \bibinfo {pages} {1} (\bibinfo {year} {2021})}\BibitemShut {NoStop}%
\bibitem [{\citenamefont {Xu}\ \emph {et~al.}(2023)\citenamefont {Xu}, \citenamefont {Huang}, \citenamefont {Fruitwala}, \citenamefont {Rajagopala}, \citenamefont {Naik}, \citenamefont {Nowrouzi}, \citenamefont {Santiago},\ and\ \citenamefont {Siddiqi}}]{xu2023qubic}%
  \BibitemOpen
  \bibfield  {author} {\bibinfo {author} {\bibfnamefont {Y.}~\bibnamefont {Xu}}, \bibinfo {author} {\bibfnamefont {G.}~\bibnamefont {Huang}}, \bibinfo {author} {\bibfnamefont {N.}~\bibnamefont {Fruitwala}}, \bibinfo {author} {\bibfnamefont {A.}~\bibnamefont {Rajagopala}}, \bibinfo {author} {\bibfnamefont {R.~K.}\ \bibnamefont {Naik}}, \bibinfo {author} {\bibfnamefont {K.}~\bibnamefont {Nowrouzi}}, \bibinfo {author} {\bibfnamefont {D.~I.}\ \bibnamefont {Santiago}}, \ and\ \bibinfo {author} {\bibfnamefont {I.}~\bibnamefont {Siddiqi}},\ }\href@noop {} {\bibfield  {journal} {\bibinfo  {journal} {arXiv preprint arXiv:2309.10333}\ } (\bibinfo {year} {2023})}\BibitemShut {NoStop}%
\bibitem [{\citenamefont {Gupta}\ \emph {et~al.}(2023)\citenamefont {Gupta}, \citenamefont {Berg}, \citenamefont {Takita}, \citenamefont {Temme},\ and\ \citenamefont {Kandala}}]{gupta2023probabilistic}%
  \BibitemOpen
  \bibfield  {author} {\bibinfo {author} {\bibfnamefont {R.~S.}\ \bibnamefont {Gupta}}, \bibinfo {author} {\bibfnamefont {E.~v.~d.}\ \bibnamefont {Berg}}, \bibinfo {author} {\bibfnamefont {M.}~\bibnamefont {Takita}}, \bibinfo {author} {\bibfnamefont {K.}~\bibnamefont {Temme}}, \ and\ \bibinfo {author} {\bibfnamefont {A.}~\bibnamefont {Kandala}},\ }\href@noop {} {\bibfield  {journal} {\bibinfo  {journal} {arXiv preprint arXiv:2310.07825}\ } (\bibinfo {year} {2023})}\BibitemShut {NoStop}%
\bibitem [{\citenamefont {Erhard}\ \emph {et~al.}(2019)\citenamefont {Erhard}, \citenamefont {Wallman}, \citenamefont {Postler}, \citenamefont {Meth}, \citenamefont {Stricker}, \citenamefont {Martinez}, \citenamefont {Schindler}, \citenamefont {Monz}, \citenamefont {Emerson},\ and\ \citenamefont {Blatt}}]{erhard2019characterizing}%
  \BibitemOpen
  \bibfield  {author} {\bibinfo {author} {\bibfnamefont {A.}~\bibnamefont {Erhard}}, \bibinfo {author} {\bibfnamefont {J.~J.}\ \bibnamefont {Wallman}}, \bibinfo {author} {\bibfnamefont {L.}~\bibnamefont {Postler}}, \bibinfo {author} {\bibfnamefont {M.}~\bibnamefont {Meth}}, \bibinfo {author} {\bibfnamefont {R.}~\bibnamefont {Stricker}}, \bibinfo {author} {\bibfnamefont {E.~A.}\ \bibnamefont {Martinez}}, \bibinfo {author} {\bibfnamefont {P.}~\bibnamefont {Schindler}}, \bibinfo {author} {\bibfnamefont {T.}~\bibnamefont {Monz}}, \bibinfo {author} {\bibfnamefont {J.}~\bibnamefont {Emerson}}, \ and\ \bibinfo {author} {\bibfnamefont {R.}~\bibnamefont {Blatt}},\ }\href@noop {} {\bibfield  {journal} {\bibinfo  {journal} {Nature communications}\ }\textbf {\bibinfo {volume} {10}},\ \bibinfo {pages} {5347} (\bibinfo {year} {2019})}\BibitemShut {NoStop}%
\bibitem [{\citenamefont {Ivashkov}\ \emph {et~al.}(2023)\citenamefont {Ivashkov}, \citenamefont {Uchehara}, \citenamefont {Jiang}, \citenamefont {Wang},\ and\ \citenamefont {Seif}}]{ivashkov2023povm}%
  \BibitemOpen
  \bibfield  {author} {\bibinfo {author} {\bibfnamefont {P.}~\bibnamefont {Ivashkov}}, \bibinfo {author} {\bibfnamefont {G.}~\bibnamefont {Uchehara}}, \bibinfo {author} {\bibfnamefont {L.}~\bibnamefont {Jiang}}, \bibinfo {author} {\bibfnamefont {D.~S.}\ \bibnamefont {Wang}}, \ and\ \bibinfo {author} {\bibfnamefont {A.}~\bibnamefont {Seif}},\ }\href@noop {} {\bibfield  {journal} {\bibinfo  {journal} {arXiv preprint arXiv:2312.14087}\ } (\bibinfo {year} {2023})}\BibitemShut {NoStop}%
\bibitem [{\citenamefont {Mallet}\ \emph {et~al.}(2009)\citenamefont {Mallet}, \citenamefont {Ong}, \citenamefont {Palacios-Laloy}, \citenamefont {Nguyen}, \citenamefont {Bertet}, \citenamefont {Vion},\ and\ \citenamefont {Esteve}}]{mallet2009single}%
  \BibitemOpen
  \bibfield  {author} {\bibinfo {author} {\bibfnamefont {F.}~\bibnamefont {Mallet}}, \bibinfo {author} {\bibfnamefont {F.~R.}\ \bibnamefont {Ong}}, \bibinfo {author} {\bibfnamefont {A.}~\bibnamefont {Palacios-Laloy}}, \bibinfo {author} {\bibfnamefont {F.}~\bibnamefont {Nguyen}}, \bibinfo {author} {\bibfnamefont {P.}~\bibnamefont {Bertet}}, \bibinfo {author} {\bibfnamefont {D.}~\bibnamefont {Vion}}, \ and\ \bibinfo {author} {\bibfnamefont {D.}~\bibnamefont {Esteve}},\ }\href@noop {} {\bibfield  {journal} {\bibinfo  {journal} {Nature Physics}\ }\textbf {\bibinfo {volume} {5}},\ \bibinfo {pages} {791} (\bibinfo {year} {2009})}\BibitemShut {NoStop}%
\bibitem [{\citenamefont {Fruitwala}\ \emph {et~al.}(2024)\citenamefont {Fruitwala}, \citenamefont {Huang}, \citenamefont {Xu}, \citenamefont {Rajagopala}, \citenamefont {Hashim}, \citenamefont {Naik}, \citenamefont {Nowrouzi}, \citenamefont {Santiago},\ and\ \citenamefont {Siddiqi}}]{fruitwala2024distributed}%
  \BibitemOpen
  \bibfield  {author} {\bibinfo {author} {\bibfnamefont {N.}~\bibnamefont {Fruitwala}}, \bibinfo {author} {\bibfnamefont {G.}~\bibnamefont {Huang}}, \bibinfo {author} {\bibfnamefont {Y.}~\bibnamefont {Xu}}, \bibinfo {author} {\bibfnamefont {A.}~\bibnamefont {Rajagopala}}, \bibinfo {author} {\bibfnamefont {A.}~\bibnamefont {Hashim}}, \bibinfo {author} {\bibfnamefont {R.~K.}\ \bibnamefont {Naik}}, \bibinfo {author} {\bibfnamefont {K.}~\bibnamefont {Nowrouzi}}, \bibinfo {author} {\bibfnamefont {D.~I.}\ \bibnamefont {Santiago}}, \ and\ \bibinfo {author} {\bibfnamefont {I.}~\bibnamefont {Siddiqi}},\ }\href@noop {} {\bibfield  {journal} {\bibinfo  {journal} {arXiv preprint arXiv:2404.15260}\ } (\bibinfo {year} {2024})}\BibitemShut {NoStop}%
\bibitem [{\citenamefont {Vora}\ \emph {et~al.}(2024)\citenamefont {Vora}, \citenamefont {Xu}, \citenamefont {Hashim}, \citenamefont {Fruitwala}, \citenamefont {Nguyen}, \citenamefont {Liao}, \citenamefont {Balewski}, \citenamefont {Rajagopala}, \citenamefont {Nowrouzi}, \citenamefont {Ji} \emph {et~al.}}]{vora2024ml}%
  \BibitemOpen
  \bibfield  {author} {\bibinfo {author} {\bibfnamefont {N.~R.}\ \bibnamefont {Vora}}, \bibinfo {author} {\bibfnamefont {Y.}~\bibnamefont {Xu}}, \bibinfo {author} {\bibfnamefont {A.}~\bibnamefont {Hashim}}, \bibinfo {author} {\bibfnamefont {N.}~\bibnamefont {Fruitwala}}, \bibinfo {author} {\bibfnamefont {H.~N.}\ \bibnamefont {Nguyen}}, \bibinfo {author} {\bibfnamefont {H.}~\bibnamefont {Liao}}, \bibinfo {author} {\bibfnamefont {J.}~\bibnamefont {Balewski}}, \bibinfo {author} {\bibfnamefont {A.}~\bibnamefont {Rajagopala}}, \bibinfo {author} {\bibfnamefont {K.}~\bibnamefont {Nowrouzi}}, \bibinfo {author} {\bibfnamefont {Q.}~\bibnamefont {Ji}},  \emph {et~al.},\ }\href@noop {} {\bibfield  {journal} {\bibinfo  {journal} {arXiv preprint arXiv:2406.18807}\ } (\bibinfo {year} {2024})}\BibitemShut {NoStop}%
\end{thebibliography}%
